\documentclass[useAMS,usenatbib]{mn2e}
\usepackage{psfig, epsf, epsfig}
\title[The star-forming environment of a ULX in NGC\,4559: an optical study]{The environment of a ULX in NGC\,4559: an optical study}
\author[Roberto Soria, Mark Cropper, Manfred Pakull, Richard Mushotzky, 
Kinwah Wu]{Roberto Soria$^{1}$\thanks{E-mail:
Roberto.Soria@mssl.ucl.ac.uk}, Mark Cropper$^{1}$, Manfred Pakull$^{2}$, 
Richard Mushotzky$^{3}$, and Kinwah Wu$^{1}$\\
$^{1}$MSSL, University College London, Holmbury St Mary, Surrey RH5 6NT, UK\\
$^{2}$Observatoire de Strasbourg, 11, rue de l'Universit\'{e}, 
		F-67000 Strasbourg, France\\
$^{3}$Laboratory for High Energy Astrophysics, NASA Goddard Space Flight Center, Greenbelt, MD 20771, USA}
\begin{document}

\date{submitted 4 May 2004}

\pagerange{\pageref{firstpage}--\pageref{lastpage}} \pubyear{2004}

\maketitle

\label{firstpage}

\begin{abstract}
We have studied the candidate optical counterparts 
and the stellar population in the star-forming complex 
around the bright ULX in the western part of the 
spiral galaxy NGC\,4559, using {\it HST}/WFPC2, 
{\it XMM-Newton}/Optical Monitor, and ground-based data.
We find that the ULX is located near  
a small group of OB stars, but is not associated 
with any massive young clusters nor with any extraordinary 
massive stars. The brightest point source 
in the {\it Chandra} error circle is consistent 
with a single blue supergiant of mass 
$\approx 20 M_{\odot}$ and age $\approx 10$ Myr. 
A few other stars are resolved inside the error circle: 
mostly blue and red supergiants with inferred masses 
$\approx 10$--$15 M_{\odot}$ and ages $\approx 20$ Myr.
This is consistent with the interpretation 
of this ULX as a black hole (BH) accreting from a high-mass donor star 
in its supergiant phase, with mass transfer occurring 
via Roche-lobe overflow. 
%more massive than typical Galactic BH candidates, 
%with a high-mass donor star in its supergiant phase, transferring matter  
%onto the more massive BH via Roche-lobe overflow.
%but still formed via normal stellar evolution.
The observed optical colors and the blue-to-red supergiant ratio 
suggest a low metal abundance for the stellar population: 
$0.2 \la Z/Z_{\odot} \la 0.4$ 
(using the Padua tracks), or $0.05 \la Z/Z_{\odot} \la 0.2$ 
(using the Geneva tracks). The age of the star-forming 
complex is $\la 30$ Myr. H$\alpha$ images show that 
this star-forming region has a ring-like appearance.
We propose that it is an expanding wave of star formation, 
triggered by an initial density perturbation, 
in a region where the gas was only marginally stable 
to gravitational collapse.
We also suggest that the most likely 
trigger was a collision with a satellite dwarf galaxy going  
through the gas-rich outer disk of NGC\,4559 less than $30$ Myr ago.
The culprit could be the dwarf galaxy visible a few arcsec 
north-west of the complex. If this is the case, this system is 
a scaled-down version of the Cartwheel galaxy.
The X-ray data favour a BH more massive ($M > 50 M_{\odot}$) 
than typical Milky Way BH candidates. The optical data 
favour a ``young'' BH originating in the recent episode 
of massive star formation; however, they also rule out 
an association with young massive star clusters (none 
are present in the X7 field). We speculate that other 
mechanisms may lead to the formation of relatively 
massive BHs (perhaps $M \sim 50$--$100 M_{\odot}$) 
from stellar evolution processes in low-metallicity 
environments, or when star formation is triggered 
by galactic collisions.

\end{abstract}


\section{Introduction}

Ultraluminous X-ray sources (ULXs) with apparent 
(isotropic) luminosities $> 10^{39}$ erg s$^{-1}$ 
have been detected in many spiral, irregular and elliptical 
galaxies; however, most sources brighter 
than $\approx 5 \times 10^{39}$ erg s$^{-1}$ 
are located in star-forming galaxies, thus suggesting that  
they are associated with young populations (Irwin et al.~2004; Swartz et al.~2004). 
The masses, ages and mechanisms of formation of the 
accreting objects are still unclear, as is the geometry 
of emission. For most of the sources, multi-band 
observations so far have not been able to rule out  
any of the three main competing ULX models, 
namely intermediate-mass black holes (Colbert \& Mushotzky 1999), 
normal X-ray binaries with beamed emission (Fabrika \& Mescheryakov~2001; 
King et al.~2001; K\"{o}rding, Falcke \& Markoff~2002), 
or with super-Eddington emission from inhomogeneous disks 
(Begelman 2002). These different scenarios also predict different 
duty cycles and timescales for the X-ray-bright phases, 
thus leading to different predictions for the total number 
of such systems (active or quiescent) in a galaxy.
However, in a few cases, such as for the brightest ULX in M\,82 
(Strohmayer \& Mushotzky 2003) and for
NGC\,4559 X7, the subject of this paper, 
there is spectral and timing evidence against beaming  
(Cropper et al.~2004; henceforth: Paper I).

The identification of an optical counterpart 
can provide important information, complementing the X-ray data.
For example, it can help in constraining the age, the mechanism 
of mass transfer and the mass-transfer rate 
in the system. X-ray ionised emission nebulae 
have been found around many ULXs (Pakull \& Mirioni 2002; 
Kaaret, Ward \& Zezas~2004): 
in some cases, they provide strong evidence against 
the beaming model (e.g., for Holmberg II X-1: Pakull \& Mirioni 2002); 
in other cases, instead, they suggest anisotropic 
emission (IC\,342 X-1: Roberts et al.~2003).
The presence of a young, massive star cluster at or near 
a ULX position (as suggested for the ULXs in the Antennae: 
Zezas et al.~2002) would be consistent with the formation 
of an intermediate-mass BH from merger processes 
in a dense cluster core (Portegies-Zwart et al.~2004).
In some cases, the identification of an optical counterpart 
has shown that the X-ray source was not a ULX but a background 
AGN (Masetti et al.~2003). In one other case, 
a strong limit on the mass of the donor star ($M < 1 M_{\odot}$) 
suggests that the supposed ULX is instead 
a foreground AM Her system (Weisskopf et al.~2004). 
However, in general, unique identifications 
of ULX optical counterparts 
have proved difficult, even when {\it HST} and {\it Chandra} 
images are available. In the few cases when reliable 
stellar identifications have been proposed (NGC\,1313 X-2: 
Zampieri et al.~2004; NGC\,5204: Roberts et al.~2001; Liu, 
Bregman \& Seitzer 2004; NGC\,3031 X-11: Liu, 
Bregman \& Seitzer 2002), the candidate donor stars 
have masses $\approx 15$--$30 M_{\odot}$.  

Located at a distance of $\approx 10$ Mpc (Sanders 2003; 
Tully 1988), the late-type spiral NGC\,4559 (Type SAB(rs)cd) 
hosts three ULXs with isotropic luminosities $\ga 10^{39}$ erg s$^{-1}$; 
two of them have X-ray luminosities $\ga 10^{40}$ erg s$^{-1}$ 
(Paper I). Its low foreground Galactic absorption 
($n_{\rm H} \approx 1.5 \times 10^{20}$ cm$^{-2}$; Dickey \& Lockman 1990)
makes it suitable for X-ray and optical studies 
aimed at determining the nature of these sources, 
and their relation with the host environment.
A study of the X-ray timing and spectral properties of the two 
brightest X-ray sources in NGC\,4559, based on {\it XMM-Newton} 
and {\it Chandra} data, was presented in Paper I; see also 
Roberts et al.~(2004). Here, we investigate  
the possible optical counterparts and stellar environment 
of NGC\,4559 X7 (adopting the naming convention 
of Vogler, Pietsch \& Bertoldi 1997 and Paper I). 
For convenience, its main X-ray properties are also summarized 
here (Table 1). A study of the optical counterparts and 
environments of the other bright X-ray sources in this galaxy 
is left to further work (HST/ACS observations scheduled for 
2005 March).

%__________________________________________________ One column table
   \begin{table}
      \caption[]{Main X-ray fitting parameters for NGC\,4559 X7, 
for an absorbed blackbody plus power-law model 
({\tt wabs$_{\rm Gal}$~$\ast$ tbvarabs $\ast$ (bb + po)} 
in {\footnotesize XSPEC}). 
The data are from the {\it Chandra} and {\it XMM-Newton} 
observations of $2001$--$2003$. 
We list here the isotropic emitted luminosity 
and the luminosity in the thermal component, 
in the $0.3$--$10$ keV band,  
%in units of $10^{40}$ erg s$^{-1}$, 
assuming $d = 10$ Mpc and $n_{\rm H,\,Gal}$~$=1.5 \times 10^{20}$ cm$^{-2}$.}
         \label{table1}
\begin{centering}
         \begin{tabular}{lccc}
            \hline
            \hline
            \noalign{\smallskip}
%            Parameter & Value & Value  
%		& Value \\
           Parameter  & 2001 Jan & 2001 Jun 
		& 2003 May\\
            \noalign{\smallskip}
	    \hline
            \hline
            \noalign{\smallskip}
            \noalign{\smallskip}
		$Z$ ($Z_{\odot}$) & $(0.31)~^{\mathrm{(a)}}$ 
			& $(0.31)~^{\mathrm{(a)}}$ & 
			$0.31^{+0.27}_{-0.20}$\\[5pt]
		$n_{\rm H}~(10^{21}~{\rm cm}^{-2})$  & $3.6^{+0.9}_{-1.1}$ & 
			$5.7^{+0.9}_{-1.1}$ &$4.3^{+0.9}_{-1.1}$\\[5pt]
		$kT_{\rm bb}$~(keV) & $0.12^{+0.06}_{-0.06}$ 
			& $0.12^{+0.01}_{-0.01}$ 
			& $0.12^{+0.01}_{-0.01}$ \\[5pt]
		$K_{\rm bb}~(10^{-5})$ & $0.9^{+0.1}_{-0.4}$ 
			& $2.3^{+0.5}_{-0.5}$ 
			& $0.8^{+0.8}_{-0.4}$ \\[5pt]
		$\Gamma$ & $1.80^{+0.08}_{-0.08}$ 
			& $2.13^{+0.08}_{-0.08}$ 
			& $2.23^{+0.06}_{-0.05}$ \\[5pt]
		$K_{\rm po}~(10^{-4})$ & $1.8^{+0.2}_{-0.2}$ 
			& $2.9^{+0.2}_{-0.2}$ 
			& $2.4^{+0.3}_{-0.2}$ \\[5pt]
		$L_{\rm x}~(10^{40}~{\rm erg~s}^{-1})$ 
			& $2.0$ & $3.4$ & $2.0$ \\[3pt]
		$L_{\rm x}^{\rm bb}~(10^{40}~{\rm erg~s}^{-1})$ 
			& $0.6$ & $1.6$ & $0.5$ \\[3pt]
%		$L_{\rm x}^{\rm po}~(10^{40}~{\rm erg~s}^{-1})$ 
%			& $2.9$ & $2.9$ & $400$ \\[3pt]
	    \hline
         \end{tabular}
\begin{list}{}{}
\item[$^{\mathrm{a}}$] Fixed from the {\it XMM-Newton} value.
%\item[$^{\mathrm{a}}$] in units of $10^{21}~{\rm cm}^{-2}$
%\item[$^{\mathrm{b}}$] in units of $10^{-12}~{\rm erg~cm}^{-2}~{\rm s}^{-1}$
%\item[$^{\mathrm{c}}$] in units of $10^{40}~{\rm erg~s}^{-1}$
\end{list}
\end{centering}
   \end{table}

\section{Optical observations}

NGC\,4559 X7 is located in a relatively isolated region 
of young star formation at the edge of the galactic disk, 
$\approx 2\farcm0$ south-west of the nucleus (Fig.~1). 
Since our viewing angle is $69^{\circ}$ (Tully 1988), this corresponds to 
a distance $\approx 16 d_{10}$ kpc, where $d_{10}$ is the distance 
to NGC\,4559 in units of $10$ Mpc. A log of the {\it Chandra} and 
{\it XMM-Newton} observations is given in Paper 1. In the optical band, 
the field of NGC\,4559 X7 was observed by {\it HST}/WFPC2 on 2001 May 25, 
with a series of four 500-s exposures in each of the three broad-band 
filters f450w ($\approx B$), f555w ($\approx V$) and f814w ($\approx I$).
We retrieved the combined WFPC2 associations as well as the individual 
exposures from the MAST public archive.

The X-ray position of X7 is located on the PC chip, 
in the WFPC2 images. The star-forming complex 
around the X-ray source is also entirely contained  
in the PC chip image (Fig.~2).
The astrometric registration of the MAST associations has errors 
$\ga 1\arcsec$ (i.e., the same stars have different WCS coordinates 
in the three combined images). In the absence of bright stars 
in the PC chip, we used stars from the USNO-B1.0 Catalog 
(Monet et al.~2003) 
to correct the relative and absolute astrometric registration 
of the three images. We estimate that we have thus reduced 
the astrometric error to $< 0\farcs4$ for the PC frame. 
We also applied this new astrometric registration to a mosaiced 
WFPC2 image, verifying that the error is $\la 0\farcs4$ 
in all chips. The {\it Chandra} astrometry is generally accurate 
to $\la 0\farcs6$. In this case, we verified 
that the position of the nuclear X-ray source coincides 
to within $\approx 0\farcs3$ with the position of one of the two 
nuclear sources in the 2MASS catalogue. Only two other 
{\it Chandra} sources have counterparts in the 
USNO-B1.0 Catalog: the differences 
between the X-ray and optical positions are 
$\approx 0\farcs4$ and $\approx 0\farcs2$. 
We can therefore take $0\farcs6$ as a conservative upper limit 
to the Chandra astrometric error, and $0\farcs7$ ($\simeq 34$ pc) 
as the total error when we overlap the X-ray and optical images.

\begin{figure}
%  \vspace*{174pt}
\psfig{figure=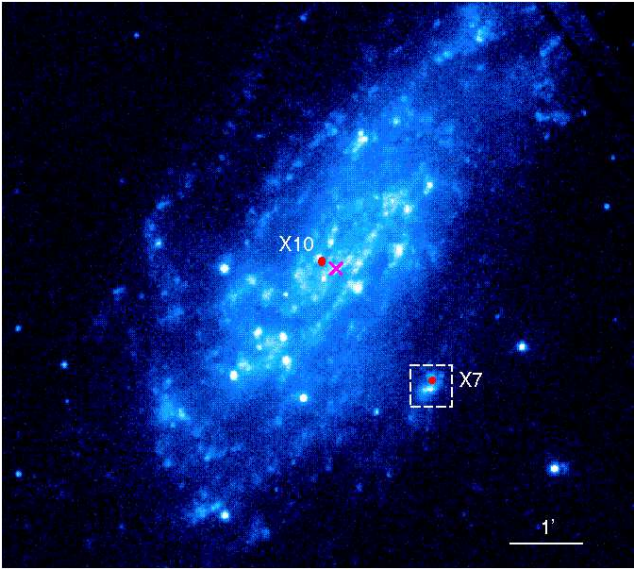,width=84mm}
  \caption{{\it XMM-Newton} Optical Monitor image of NGC\,4559, 
in the UVW1 filter. North is up, east is to the left. 
The locations of the two brightest ($L_{\rm x} > 10^{40}$ erg s$^{-1}$) 
X-ray sources are marked with red circles, and the galactic nucleus 
(also an X-ray source) with a magenta cross.  The area inside 
the dashed square is shown in more details in Fig.~2, 
from {\it HST}/WFPC2.}
\end{figure}

\begin{figure}
%  \vspace*{174pt}
\psfig{figure=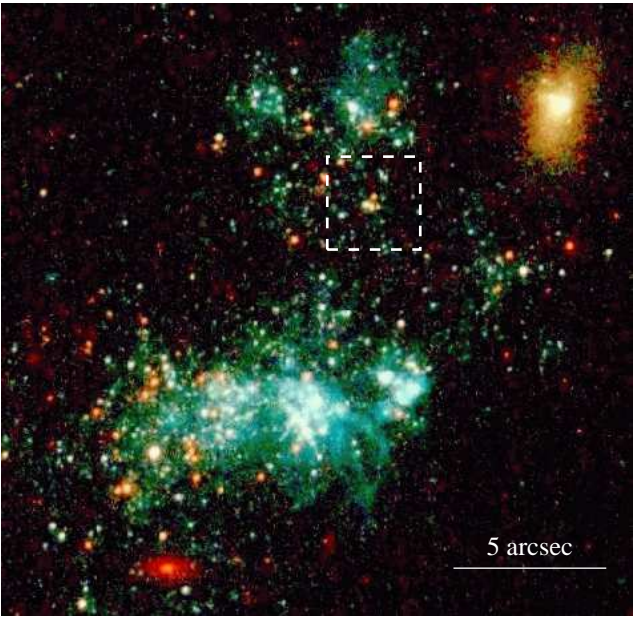,width=84mm}
  \caption{{\it HST}/WFPC2 (PC chip) true-color image 
of the field around NGC\,4559 X7. North is up, east is to the left. 
Red, green and blue colours correspond to the f814w, f555w and f450w 
filters respectively. Notice the color difference 
between the star-forming complex, populated by young massive stars, 
and the old (yellow) population in the galaxy near the top right corner.
The area inside the dashed square is shown in more details in Fig.~3. }
\end{figure}

\begin{figure}
%  \vspace*{174pt}
\psfig{figure=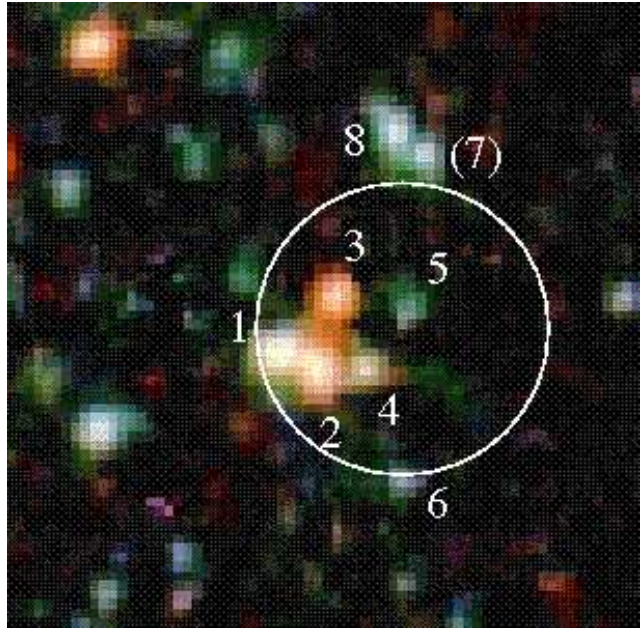,width=84mm}
  \caption{Close-up view of the immediate environment 
of X7. The ULX is located inside  
a circle with $0\farcs7$ radius (this combines 
the error in the {\it Chandra} pointing accuracy and 
that in the {\it HST}/WFPC2 astrometry).
The brightness of the labelled stars around X7 
%(identified by their respective labels)
is listed in Table 2.}
\end{figure}

%\section{Masses, ages and metallicity of the stellar population}

%\subsection{Stellar population in the ULX field}
\section{Data analysis}

Before carrying out the photometric analysis, 
we corrected the WFPC2 images for geometric distortions, 
multiplying the image files by the standard correction file 
({\tt f1k155bu.r9h}) available from the WFPC2 
web site\footnote{http://www.stsci.edu/instruments/wfpc2}
(Gonzaga 2002).
We then used {\tt daophot} tasks in {\footnotesize IRAF} 
to detect stellar sources and measure their brightness.

We have performed a detailed photometric analysis 
for the sources in the PC chip ($1.6 \times 1.6$ kpc$^2$ 
field of view at the distance of NGC\,4559), 
which includes the whole star-forming complex around X7.
We detected $\approx 260$ stars or star-like objects 
(after rejecting a few extended objects, likely 
to be background galaxies), 
at the $4\sigma$ level in the f555w-filter image.
We also ran the source-finding 
routine {\tt {daofind}} independently on the f450w and f814w images, 
and compared the three lists. This did not add any 
new stellar objects, i.e., there are no stars detected 
at the $4\sigma$ level in either filter that 
are not also detected in the f555w filter.
The significance of detection is, in general, slightly higher 
in the f555w filter than in the other two; in other words, some 
of the stars detected at the $4\sigma$ level 
in f555w are only $\approx 3\sigma$ detections 
in f450w and f814w. 
Setting the threshold at $3.5\sigma$ in the f555w filter 
or using a combined image in the source-finding routine 
led to the detection of more stars, 
but we verified that the additional sources were  
too faint for any meaningful color analysis.

We applied the $4\sigma$-detection source list 
to each filter image, and used {\tt phot} and {\tt allstars} 
to obtain instrumental magnitudes for the point sources 
in all three filters. Having to deal 
with a crowded field, we built a point spread function 
from the brightest, isolated stars, and carried out profile-fit 
photometry rather than simple aperture photometry.
The zeropoints were taken from Andrew Dolphin's 
up-to-date tables\footnote{http://www.noao.edu/staff/dolphin/wfpc2\_calib/} 
(last updated: 2002 Sept 17; see also Dolphin 2000; Holtzman et al. 1995).
We also applied A. Dolphin's equation 
to carry out the charge-transfer-efficiency corrections.
We then applied the colour transformation 
coefficients listed in A. Dolphin's website to obtain 
standard $B, V, I$ colours in addition to 
the instrumental magnitudes.
Finally, we added and tried to detect artificial stars 
in our images, to determine the completeness limit. 
We estimate that our sample 
is complete down to $m_{555} \approx V \approx 24.9$.
%\vspace{1cm}

Colour-magnitude diagrams of the stars in the X7 field 
allowed us to study their age and mass distribution.
We compared the observed colours with two sets 
of evolutionary tracks and isochrones for single stars: 
the Geneva models (Lejeune \& Schaerer 2001; 
Meynet et al.~1994; data files available on VizieR), 
and the Padua models (Fagotto et al.~1994; Girardi et al.~2002; 
Salasnich et al.~2000; data files available on VizieR 
and at http://pleiadi.pd.astro.it/).
For the Geneva models, we used the ``e'' type of grid 
(enhanced mass loss), which Lejeune \& Schaerer (2001) 
suggest to be preferable for stars of mass $> 7 M_{\odot}$.
For the Padua models, we used solar-scale rather 
than $\alpha$-enhanced grids (see Salasnich et al.~2000 
for a discussion of this issue).
Whenever possible, we compared the observed colors 
directly with the tracks and isochrones in the 
WFPC2 photometric system; such models are available 
for f555w and f814w. However, they are not available 
for the blue filter f450w: the other, more commonly used blue 
filter f439w is generally chosen for synthetic models, because 
it is a closer approximation to the standard $B$ filter. 
In that case, we compared the $(B, B-V)$ model tracks 
with the $(B, B-V)$ colors of the observed stars, 
obtained with Dolphin's transformation coefficients, 
as mentioned above. To convert the theoretical stellar 
tracks from absolute to apparent magnitudes, 
we assumed a distance modulus $d_{\rm M} = 30.0$ mag 
(corresponding to a distance of $10$ Mpc) throughout our analysis. 
We also corrected the theoretical tracks for the line-of-sight 
Galactic extinction $E(B-V)= 0.018$ (Schlegel, Finkbeiner \& 
Davis~1998).
 
We are aware that all the available sets of evolutionary 
tracks and isochrones are calculated for single stars, 
while most O stars come in double or triple systems. 
Besides, some of the bright datapoints in our color-magnitude 
diagrams may in fact refer to a superposition of two or more 
fainter, unresolved stars. These are well known 
limitations of stellar population studies. Nonetheless, 
similar studies carried out for nearby galaxies 
have generally given a reliable view of the stellar population 
properties.

\begin{figure*}
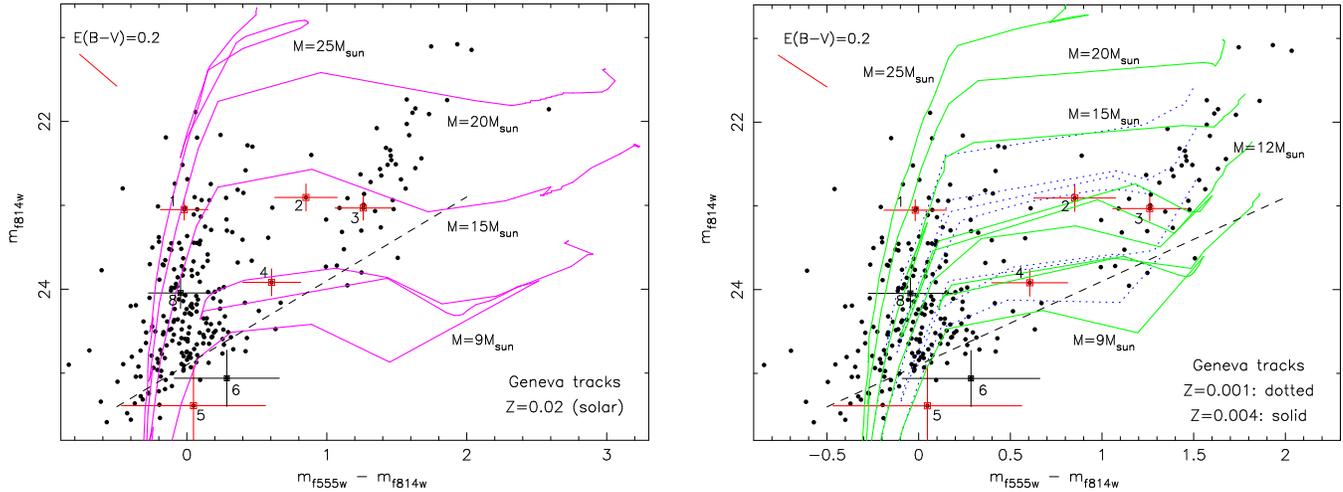

%  \vspace*{174pt}
\epsfig{file=cmd1genS.ps,width=6.45cm, angle=270}
\hspace{0.5cm}
\epsfig{file=cmd1gen3.ps,width=6.45cm, angle=270}
  \caption{Color-magnitude diagrams 
of the stars in the young complex around X7 (the region 
shown in Fig.~2), in the (f555w,\,f814w) WFPC2
	filter system ($\approx (V,I)$), and 
	evolutionary tracks for single stars of various 
initial masses and metallicities (Geneva tracks). 
 The stars plotted with error bars 
and number labels are the candidate optical counterparts 
of the ULX: see Table 2, Fig.~3 and Sect.~4.4. 
The tracks have already been corrected for line-of-sight reddening.
The bar near the top left corner shows 
the effect on the data of an (arbitrary) intrinsic 
reddening $E(B-V) = 0.2$. The dashed black line 
is the completeness limit of our sample.
Left panel: solar abundances are ruled out by the small 
colour separation observed between the MS/BSG clump and the RSG clump 
(only $(V-I) \approx 1.5$ mag, instead of $(V-I) \approx 2.5$ mag 
as expected for solar metallicity).
Right panel: the Geneva tracks providing the best model 
for the observed stellar distribution are those 
for metal abundances $0.001 \la Z \la 0.004$, 
consistent with the low abundance inferred from 
the X-ray data. In this panel, we have plotted 
evolutionary tracks for initial masses of $9, 12$ and $15 M_{\odot}$ 
at $Z = 0.001$ (dotted blue lines), and 
tracks for initial masses of $9, 12, 15, 20$ and $25 M_{\odot}$ 
at $Z = 0.004$ (solid green lines).}
\end{figure*}

\begin{figure*}
%  \vspace*{174pt}
\epsfig{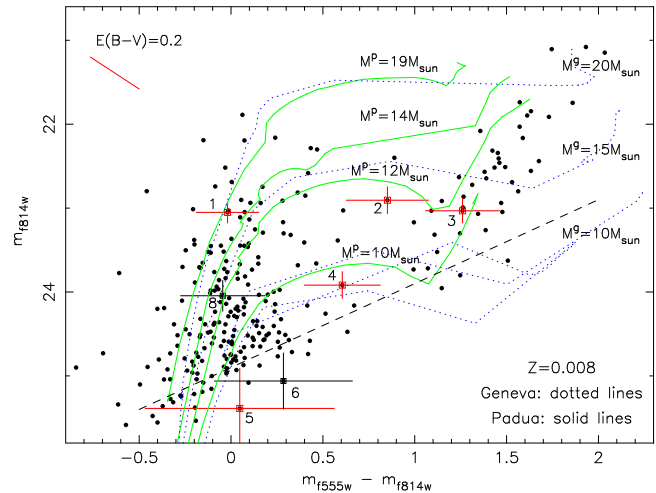}
  \caption{Color-magnitude diagram in the (f555w,\,f814w) WFPC2
	filter system, illustrating the comparison between 
Geneva and Padua tracks. At a fixed metal abundance, 
tracks from the Geneva models (dotted blue lines) 
are systematically redder than tracks from the Padua model 
(solid green lines). Error bars and symbols are the same as in Fig.~4.}
\end{figure*}

\begin{figure*}
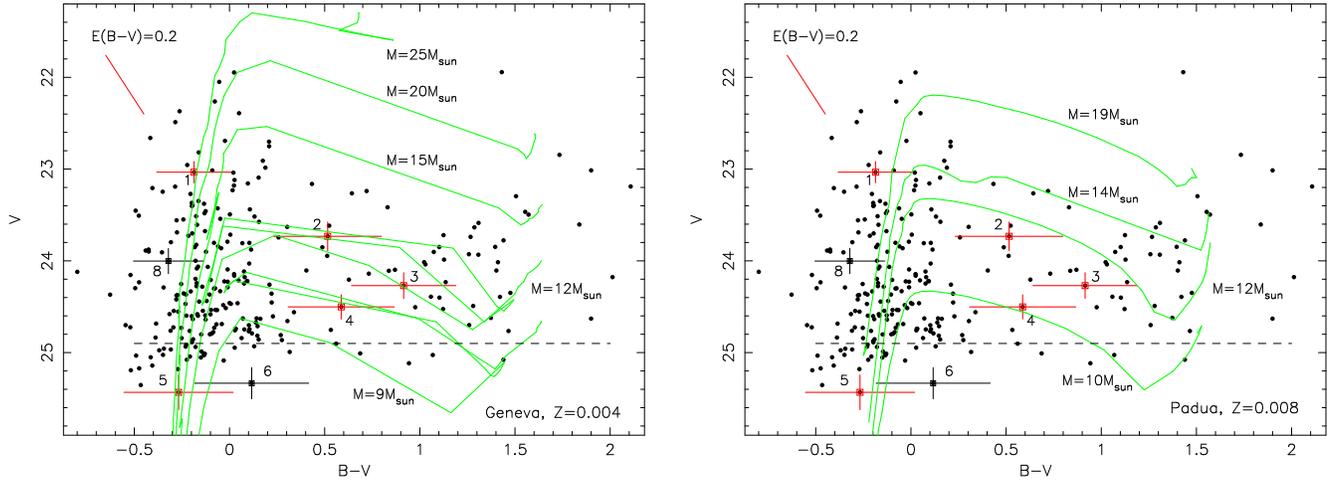

%  \vspace*{174pt}
\epsfig{file=cmd2gen.ps,width=6.3cm, angle=270}
\hspace{0.4cm}
\epsfig{file=cmd2padua.ps,width=6.3cm, angle=270}
  \caption{Color-magnitude diagrams 
of the stars in the X7 field, in the standard $(B,\,V)$ system, 
and evolutionary tracks for single stars of various 
initial masses. Left panel: tracks from the Geneva models; 
right panel: tracks from the Padua models. Error bars and labels 
have the same meaning as in Fig.~4.}
\end{figure*}

\section{Stellar population in the ULX field}

\subsection{Overview}

We divide our analysis into two parts.
Firstly (Sects. 4.2, 4.3), we shall discuss the general 
properties of the stellar population in the star-forming 
complex (metal abundance, characteristic masses and ages).
Then (Sect. 4.4, 4.5), we shall focus on the brightest stars detected 
within $1\arcsec$ from the X-ray position of the X7 ULX, 
discussing whether any of them can be the donor stars 
for the accreting BH. 

The brightness in the standard system of the candidate 
X7 counterparts is listed in Table 2; 
the numbers correspond to those overplotted 
in Fig.~3. The same set of stars has been plotted 
with their respective error bars in all color-magnitude diagrams 
throughout this Section (Figs.~4, 5, 6, 9, 10), 
to be more easily identified 
and to provide an indication of typical photometric errors 
at various brightnesses and colors.

\subsection{Metal abundance}

The first apparent result (independent of the choice of comparison 
models) from the observed optical color distribution 
of the stars from the whole star-forming complex is the clear 
separation between two groups of blue and red stars (Fig.~4). 
The latter ($m_{555}-m_{814} \ga 0.7$) are easily identified 
as red supergiants (RSG: stars with shell Hydrogen 
or, later, shell Helium burning); we find 46 of them. 
The blue stars comprise blue supergiants 
(BSG: stars in their core-Helium-burning phase of evolution), 
and massive stars ($M \ga 15 M_{\odot}$) slightly evolved 
but still on the main sequence (MS). 
An unequivocal separation between these two classes 
of bright blue stars in a stellar population 
is usually difficult (Dohm-Palmer \& Skillman 2002).
Using a color-index histogram of Sextans A stars 
with absolute magnitude $-5.6 < M_V < -3.2$, 
Dohm-Palmer \& Skillman (2002) showed that, 
although there is some overlapping, most 
of the stars with $(V-I)_0 < -0.13$ are MS stars, 
and most of those with $(V-I)_0 > -0.13$ 
are BSG. This dividing line is independent of $M_V$.
They also found approximately the same relative 
contribution from the two populations in that magnitude range.

NGC\,4559 is 7 times further away than Sextans A, 
so our detection limit is at $M_V \approx -5$, 
and there is a higher chance of confusion. 
At magnitudes $M_V < -5$, separating 
between the two subgroups of blue stars is less 
straightforward (see Fig.~2 in Dohm-Palmer \& Skillman 2002). 
%from all stellar evolution models, we also 
%expect a larger relative contribution from BSG 
%at brighter magnitude ranges.
If we assume that the dividing line found by 
Dohm-Palmer \& Skillman (2002) also applies 
to our sample, i.e., if we identify as MS all the stars 
bluer than $(V-I)_0 = -0.13$, 
corresponding to an observed color index 
$m_{555} - m_{814} \approx -0.10$, we obtain (Fig.~7) 
that $\approx 2/3$ of the blue stars 
in the X7 field are BSG (i.e., $\approx 140$).
We can conservatively conclude that 
$\approx 100$--$150$ stars are BSG 
and $\approx 40$--$50$ are RSG.

Hence, we estimate a blue-to-red supergiant ratio $\approx 3\pm1$, 
similar to the values observed in metal-poor galaxies 
such as the SMC (Langer \& Maeder 1995) or Sextans A
(Dohm-Palmer \& Skillman 2002). This ratio 
is one order of magnitude higher in metal-rich 
environments, at solar abundance or higher (Langer \& Maeder 1995). 
The metal dependence in the evolutionary tracks of massive stars 
has been explained (Maeder \& Meynet 2001) 
as an effect of rotation: metal-poor stars lose 
less mass and angular momentum in a wind, are more likely 
to be fast rotators (close to their break-up speed), 
end their H-burning phase with higher He-core masses 
and expand to much larger radii.
(For recent studies and reviews of the blue-to-red 
supergiant ratio, see Langer \& Maeder 1995; Salasnich, 
Bressan, \& Chiosi 1999; Maeder \& Meynet 2001; Guo \& Li 2002). 

Another indication of the metal-poor nature of 
the stellar population in the X7 field comes directly 
from the colors of the RSG population (Figs.~4, 5, 6).
For a fixed metal abundance, the color indices  
predicted by the Geneva and Padua tracks 
differ by $\sim 0.5$--$1$ mag (Fig.~5), 
with the Geneva tracks systematically redder. 
However, both sets of models indicate 
a subsolar metallicity: $0.2 \la Z/Z_{\odot} \la 0.4$ 
(Padua), or $0.05 \la Z/Z_{\odot} \la 0.2$ (Geneva).

In conclusion, both the blue-to-red supergiant ratio and 
the RSG color indices point to a low metal abundance, 
similar to values found in the SMC and in other 
nearby dwarf galaxies. An association between ULXs 
and metal-poor environments was suggested 
by Pakull \& Mirioni (2002). 
This is also consistent with the abundance 
$Z/Z_{\odot} = 0.3^{+0.3}_{-0.2}$ inferred from 
the X-ray spectral fits to the X7 ULX (Table 1 and Paper I).

\begin{figure}
%  \vspace*{174pt}
\epsfig{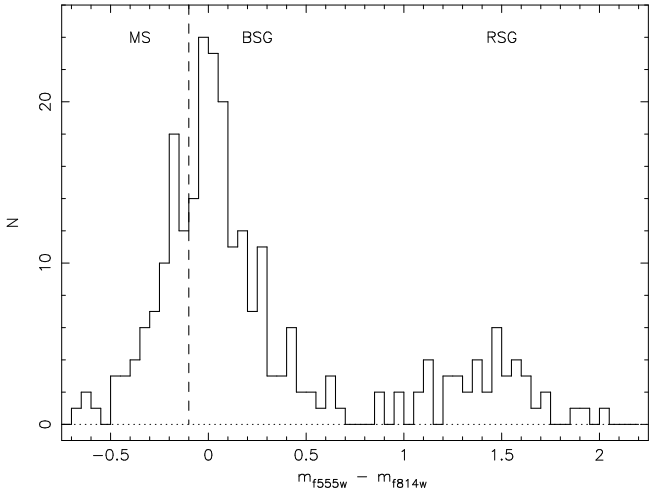}
  \caption{Color histogram of the stars in the X7 field, showing 
two separate groups of blue (MS $=$ main sequence, and BSG $=$ 
blue supergiants) and red stars (RSG $=$ red supergiants). 
We cannot separate the MS and BSG populations; however, 
following Dohm-Palmer \& Skillman (2002), we estimate 
that the stars to the left of the vertical dashed line 
are more likely to be MS, and those to the right BSG.}
\end{figure}

\begin{figure}
%  \vspace*{174pt}
\epsfig{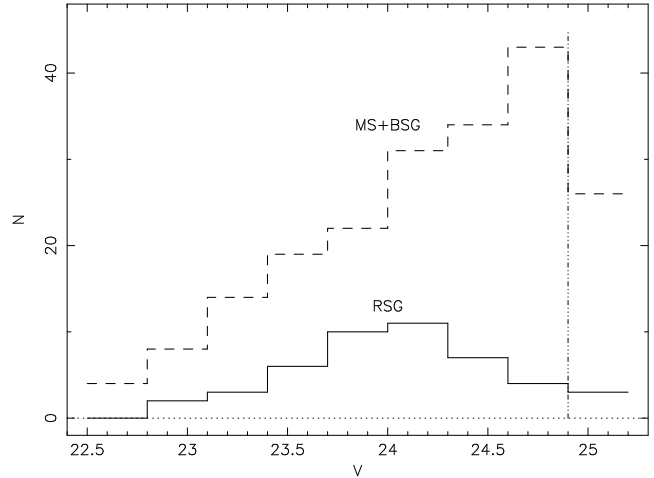}
  \caption{$V$ brightness histogram of the stars 
in the X7 field, showing that most RSG are located 
at $V \approx 24$. From this, we infer a characteristic 
age for the recent star formation in the region (see text).
The vertical dash-dotted line is the completeness limit.}
\end{figure}

\subsection{Masses and ages}

Various differences between the Geneva and Padua models 
have been extensively discussed in the literature 
(e.g., Westera et al.~2002; 
Lejeune \& Buser 1999; Fioc \& Rocca-Volmerange 1997). 
It is well known that, for a given metal abundance, 
the Geneva tracks for the red giant 
and supergiant phases tend to be redder than observed, 
and the Padua tracks tend to be bluer, especially at low 
metallicities; or alternatively, that the Geneva tracks 
predict systematically lower metal abundances than 
the Padua tracks. This is also evident for example from Fig.~5.

For the stellar population in the X7 field, 
both the Geneva and the Padua tracks suggest that 
most of the RSG have masses $\approx 12$--$14 M_{\odot}$ (Figs.~4, 5, 6), 
with only a few more massive ones, up to $M \approx 20 M_{\odot}$.
On the blue side of the colour-magnitude diagrams, 
there is evidence for stars 
with masses $\approx 25 M_{\odot}$ near the end of 
their MS phase or in the BSG stage.

The largest number of RSG are located at $23.9 \la V \la 24.3$ (Fig.~8), 
corresponding to stars with initial masses $\approx 12 M_{\odot}$. 
A decreasing number of RSG is found at fainter $V$ brightnesses 
(corresponding to lower masses).  
This is not an effect of incompleteness, because 
our source list is complete down to $V \approx 24.9$. 
Indeed, the number of detected blue stars (MS plus BSG) 
increases at fainter brightness intervals, down 
to the completeness limit, as expected (Fig.~8). 
The peak in the RSG distribution at $V \approx 24.1$ is 
likely due to the young age of the population: stars 
with masses $\la 10 M_{\odot}$ have not yet evolved 
to the RSG stage.

We estimated the dominant age of the population by overplotting 
the Geneva and Padua theoretical isochrones, 
for various metal abundances (Figs.~9, 10).
%As expected from the evolutionary tracks analysis 
%(Sect.~4.1), the observed colour indices require 
%metal abundances lower than $Z_{\odot}$ for both sets 
%of models, consistently with the evolutionary tracks
We find that the median age of the stars is 
$\approx 15$--$20$ Myr, depending on the choice 
of model and metal abundance. A few bright, 
blue stars could be as young as $\approx 3$ Myr. 
The brightness distribution of the RSG 
suggests that star formation was much reduced 
at ages $\ga 30$ Myr. In fact, the integrated colour 
of the whole star-forming complex (rather blue, with 
$B - I \approx 0.0$) suggests that, if there is an older, underlying 
stellar population, its contribution is negligible.

A visual inspection of the stellar colors 
in Fig.~2 suggests that the main star-forming region 
south of the X7 ULX does not have a uniform population: 
most of the RSG are located on the eastern side, 
and most of the blue stars and gas are on the western side.
To estimate a possible age gradient more quantitatively, 
we divided this region into two halves: an eastern and 
a western region containing 88 and 89 stars respectively. 
Overplotting the Geneva isochrones 
onto their colour-magnitude diagram (Fig.~11, left panel), we find 
that the stellar population in the eastern part of the complex 
is consistent with a narrow age range ($\approx 16$--$18$ Myr), 
with only a few stars as old as $\approx 30$ Myr 
and no stars younger than $\approx 10$ Myr.
In the western region, instead, 
we find both young, bright BSG and MS stars 
with an age $\approx 3$--$10$ Myr, 
and fainter, blue stars, apparently BSG older 
than $\approx 25$ Myr and with masses $\la 10 M_{\odot}$. 
The large error in their colors prevents an accurate 
age determination for this fainter group. It is puzzling to find 
so many BSG in that age and mass range without a corresponding 
number of RSG. However, these stars are mostly located 
in a gas-rich area, where we expect a local extinction 
larger than the Galactic line-of-sight value. 
We think that these apparently older stars 
are in fact young but simply more reddened. 
Sample reddening bars are plotted in the top left 
corner of all colour-magnitude diagrams (Figs.~4--6, 9--12).

Star formation in the X7 field probably 
proceeded through successive bursts, igniting 
different parts of the complex at different times.
In particular, over the last $20$ Myr, star formation seems 
to have propagated from the south-eastern 
corner of the complex towards the west.
A similar analysis for the rest of the X7 field 
(including the smaller star-forming complex 
north of the ULX) reveals an age distribution 
$\approx 10$--$20$ Myr (Fig.~11, right panel), intermediate between 
those of the two halves of the southern complex. 

There is only one moderately large star cluster 
in the field, located at {R.A.~(2000) $= 12^h$\,$35^m$\,$52^s.32$}, 
{Dec.~(2000) $= +27^{\circ}$\,$55$\arcmin\,$55$\farcs{1}}, 
i.e., $\approx 12$\arcsec~($\approx 570$ pc) south-east of the ULX.
It is unresolved in the PC images (full-width half maximum 
of its radial profile $\approx 1.7$ 
pixels, similar to all other point sources in the field), 
hence its size is $\la 3.5$ pc. 
Its apparent brightness is $B = (20.65 \pm 0.14)$, 
$V = (20.49 \pm 0.10)$, $I = (19.96 \pm 0.11)$, 
hence $M_V \approx -9.5$: we have of course excluded it 
from our color-magnitude diagrams. From its optical colours 
and brightness, using the Starburst99 tables (Leitherer et al.~1999) 
for the case of instantaneous burst, Salpeter IMF, lower 
stellar-mass limit of $1 M_{\odot}$ and upper stellar-mass 
limit of $100 M_{\odot}$, we infer an age of $25\pm5$ Myr 
and a mass $\approx 10^{4.2} M_{\odot}$.
%The comparatively old age is consistent with the location 
%of this cluster near the south-eastern edge of 
%the star-forming complex. 
Such a cluster would have contained 
$\approx 70$ O stars at its birth, and would have today 
a supernova rate of $\approx 1$--$2 \times 10^{-5}$ yr$^{-1}$.
If we assumed instead a lower mass limit of $0.1 M_{\odot}$, 
the total stellar mass in the cluster would be 
$\approx 10^{4.7} M_{\odot}$.
 
The mass threshold below which the IMF flattens or turns over 
is still a matter of debate: it is generally estimated 
that the IMF flattens for $M \la 0.5 M_{\odot}$ (Kroupa 2002a; 
Kroupa, Tout \& Gilmore 1990, 1993), 
and probably at higher masses 
in low-metallicity environments such 
as the Magellanic Clouds (Larson 1998; Kroupa 2001, 2002b).
We note that young star clusters with a top-heavy IMF 
(truncated at a lower mass $\approx 1$--$3 M_{\odot}$) 
have been observed in the starburst galaxy M\,82, 
where star formation was probably triggered by tidal interactions
(Smith \& Gallagher 2001; McCrady, Gilbert \& Graham 2003; 
see also Elmegreen \& Shadmehri 2003).
Hence, we shall take $M = 1 M_{\odot}$ 
as an order-of-magnitude approximation 
for the lower cut-off mass.

\begin{figure*}
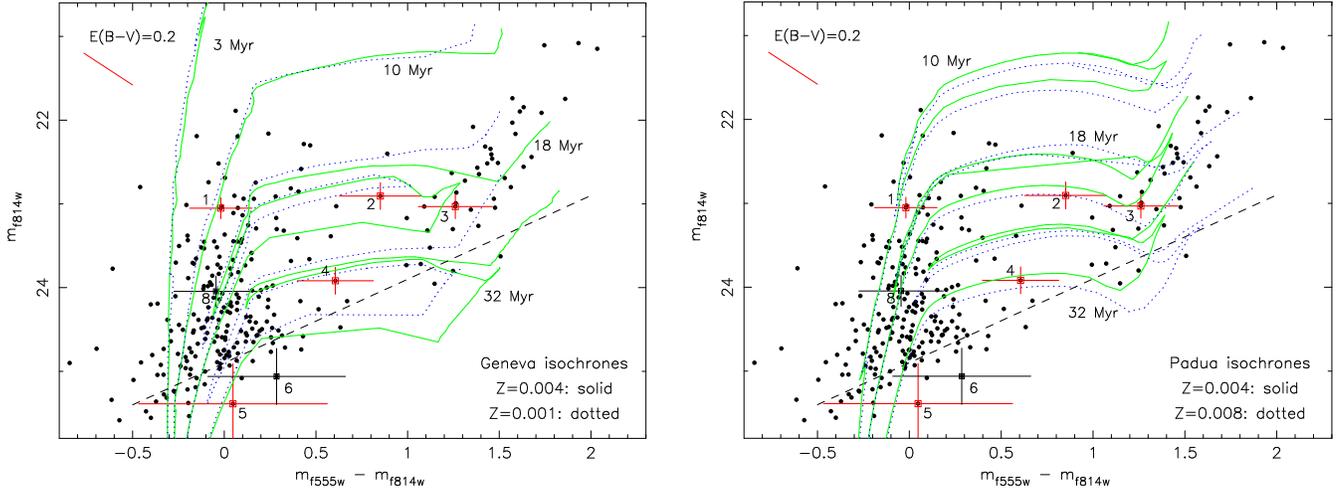

%  \vspace*{174pt}
\epsfig{file=cmd1iso.ps,width=6.45cm, angle=270}
\hspace{0.4cm}
\epsfig{file=cmd1iso_pd.ps,width=6.45cm, angle=270}
  \caption{Theoretical isochrones in the (f555w$-$f814w,\,f814w) plane, 
for different choices of metal abundances. Left panels: Geneva 
models; right panel: Padua models. The labels and error bars  
have the same meaning as in Fig.~4.}
\end{figure*}

\begin{figure*}
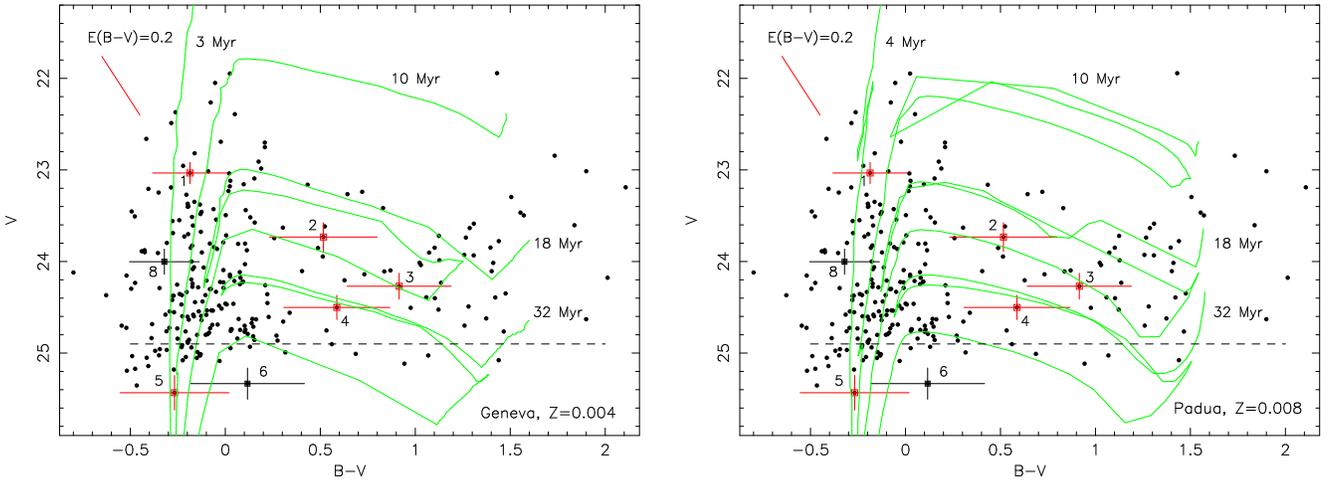

%  \vspace*{174pt}
\epsfig{file=cmd2iso_gen.ps,width=6.3cm, angle=270}
\hspace{0.4cm}
\epsfig{file=cmd2iso_padua.ps,width=6.3cm, angle=270}
  \caption{Theoretical isochrones in the $(B-V,\,V)$ plane. 
Left panel: Geneva models for $Z=0.004$; right panel: Padua models 
for $Z=0.008$. Symbols and labels as in Fig.~4.}
\end{figure*}

\begin{figure*}
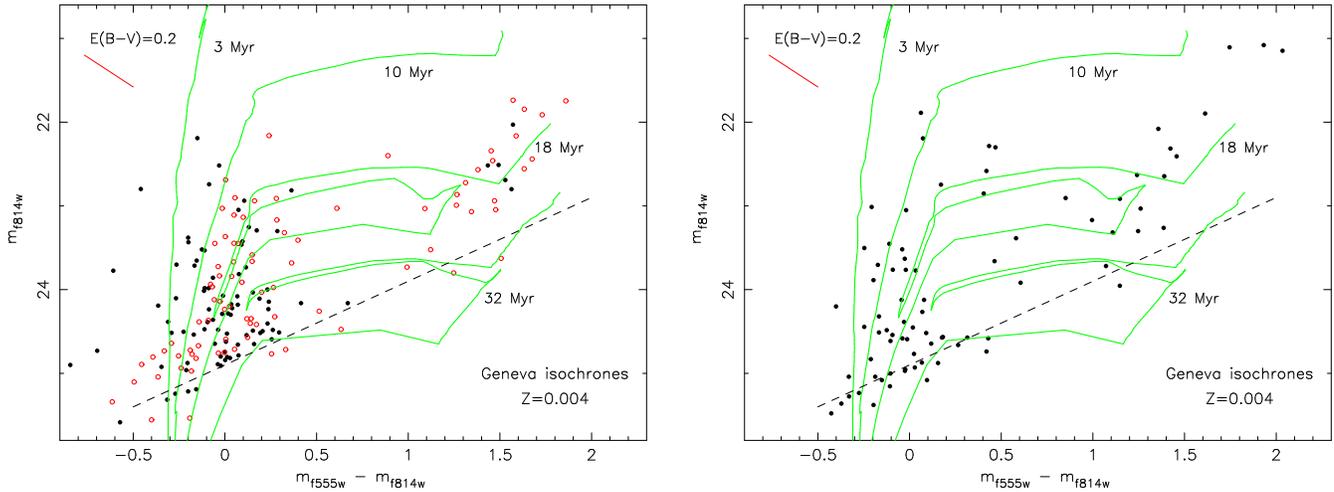

%  \vspace*{174pt}
\epsfig{file=cmd1iso_r1.ps,width=6.45cm, angle=270}
\hspace{0.4cm}
\epsfig{file=cmd1iso_r3.ps,width=6.45cm, angle=270}
  \caption{Color magnitude diagrams for three separate 
regions in the X7 field, showing age differences in the dominant 
population (also apparent from the different colours in Fig.~2). 
Left panel: the south-eastern sector of the star-forming 
complex shows a narrow age range around $18$ Mpc (open circles), 
with no stars younger than $\approx 10$ Mpc; 
the south-western sector has a younger population (filled circles), 
with ongoing star formation and fewer RSG.
Right panel: stellar population in the rest of the X7 field (including, 
among others, the candidate optical counterparts in the ULX error circle), 
showing an intermediate age spread, up to an age $\approx 25$ Myr.}
\end{figure*}

\begin{table}
% \centering
% \begin{minipage}{140mm}
  \caption{Standard magnitudes of the brightest stars 
detected within $1\arcsec$ from the ULX position.}
\centering
  \begin{tabular}{lccc}
  \hline
   No.     &   $B$ brightness &   $V$ brightness &   $I$ brightness \\
 \hline
1	& $22.85\pm0.16$  &    $23.03\pm0.12$  &  $23.05\pm0.12$ \\
2	& $24.25\pm0.24$  &    $23.73\pm0.16$  &  $22.87\pm0.16$ \\
3	& $25.18\pm0.24$  &    $24.27\pm0.14$  &  $22.99\pm0.14$ \\
4	& $25.09\pm0.25$  &    $24.50\pm0.13$  &  $23.89\pm0.16$\\
5	& $25.17\pm0.22$  &    $25.43\pm0.19$  &  $25.38\pm0.48$\\
6	& $25.45\pm0.25$  &    $25.33\pm0.17$  &  $25.05\pm0.33$\\
7	& \multicolumn{3}{c}{(falling on a hot pixel)}\\
8	& $23.68\pm0.12$  &    $24.00\pm0.14$  &  $24.05\pm0.18$\\
\hline
\end{tabular}
%\end{minipage}
\end{table}

\subsection{Nature of the donor star}

After discussing the general properties of the starburst complex, 
we now focus on the immediate environment of the ULX.
Four point-like sources are detected with signal-to-noise $> 4\sigma$ 
in the f555w-filter image, inside a $0\farcs7$ error circle 
centred on the {\it Chandra} position for X7 (Fig.~3);  
a fifth star is detected with signal-to-noise $\approx 3.5\sigma$.
We also considered three other bright stars just outside 
the error circle, at a distance $d < 1\arcsec$~from the {\it Chandra} 
position. See Fig.~3 and Table 2 for the location and 
brightness of these stars.
%The brightness in the standard system 
%of all these candidate counterparts 
%is listed in Table 2; the numbers correspond 
%to those overplotted in Fig.~3. 
(One of them, star No.~7, 
falls onto a hot WFPC2 pixel\footnote{For a list of hot {\it HST}/WFPC2 pixels 
at various epochs, and a general discussion of this problem, 
see http://www.stsci.edu/instruments/wfpc2/wfpc2\_hotpix.html.}: 
it was not possible to derive an accurate brightness.)
In our colour-magnitude diagrams (Figs.~4--6, 9, 10, 12), 
the five stars plotted with red error bars are 
those located at $< 0\farcs7$ from the X-ray position 
of the ULX, while the two stars with black error bars 
are those located at $0\farcs7 < d < 1\arcsec$.

There are clearly many more lower-mass stars in the same region 
and immediately outside the error circle (Fig.~3), 
but they are either too faint, or too close to much brighter 
stars to allow for a meaningful photometric study.
The young age of the population implies 
that all B3 or later-type stars ($M \la 9 M_{\odot}$) 
are still on the main sequence.
We assume for this work that the true ULX counterpart 
is one the more massive stars.
However, we cannot rule out the possibility that the true ULX counterpart
is one of the unresolved lower-mass stars, transferring mass 
via Roche lobe overflow on their nuclear evolution 
timescale. 
%We shall address this issue in further work.

Among the five stars resolved inside the $0\farcs7$ error circle,
star No.~1 (Fig.~3) is the brightest in the $V$ band. 
From the Geneva models, its colours are consistent with 
a zero-age mass of $\approx 15$--$30 M_{\odot}$ (Figs.~4--6), 
and an apparent age of $\approx 10$ Myr (Figs.~9, 10); it is likely 
to be in its BSG phase. If this is the case, we infer a bolometric 
luminosity $\approx 1.4\pm0.2 \times 10^5 L_{\odot}$ 
and an effective temperature $\approx 16,000\pm 5,000$ 
K\footnote{Hence, we note only as an aside that this star 
is similar to the BSG progenitor of SN 1987A; 
Podsiadlowski (1992).}.
%Podsiadlowski, P. 1992, PASP, 104, 717

The three stars labelled as No.~2, 3 and 4 (Fig.~3) have 
initial masses $\approx 10$--$12 M_{\odot}$, 
typical of early-type stars (B0--B2 types when on the main sequence); 
with an age of $\approx 20$--$30$ Myr, they have 
already left the main sequence. 
One is clearly an RSG, another is a BSG, the third one may be 
on the ``blue loop'' between the red and blue supergiant phases.
Because of the large uncertainty in the photometry, 
we cannot constrain the age and mass of star No.~5.

The bright stars around the ULX position 
appear to be part of a single group; hence, 
it is possible that they were formed at approximately the same epoch. 
We find that the Geneva isochrone for an age of $22$ Myr 
and a metal abundance $Z = 0.004$ is consistent (within $1 \sigma$) 
with six of the seven stars (Fig.~12). 
If we impose this strict age condition, we can also obtain 
a strong constraint on their masses: they would all 
be in a narrow range between $\approx 10.5$ and $11 M_{\odot}$, 
with star No.~5 being the least massive.
Only star No.~1 appears too blue and too bright 
(and therefore too massive) to be consistent 
with an age of $22$ Myr. This star is also 
our strongest candidate for the X7 donor star, 
given its high mass loss rate in its BSG stage.
If it is indeed the donor star, and if 
we assume that it was also formed at the same epoch 
as the other stars in the group, we speculate 
by analogy with typical high-mass X-ray binaries 
(e.g., LMC X-3: Soria et al.~2001) 
that its blue colors may be the result 
of X-ray irradiation. 

The fraction of radiation intercepted 
by the companion star in an X-ray binary is 
$\approx (R_2/2a)^2$, where $R_2$ 
is the radius of the star, and the binary separation 
$a = 2.9 \times 10^{11} \, M_1^{1/3} \, (1+q)^{1/3}\, P_{\rm d}^{2/3}$ cm  
(Newton 1687). Here $M_1$ is the BH mass 
in solar units, $q = M_2/M_1$ and $P_{\rm d}$ is the binary 
period in days.
If the donor star is filling its Roche lobe, or is very 
close to filling it, we can approximate $R_2/a \approx 
r_{\rm L}(q) = (0.49q^{2/3})/\left[0.6q^{2/3}+\ln(1+q^{1/3})\right]$ 
(Eggleton 1983). Taking $M_1 \approx 100 M_{\odot}$ 
and $M_2 \approx 12$--$25 M_{\odot}$, and assuming 
isotropic emission, the intercepted X-ray luminosity is 
$\sim 2$--$6 \times 10^{38}$ erg s$^{-1}$ (Paper I),
to be partly reflected and partly thermalised 
and re-radiated in the optical/UV bands.
%%%%%%%%%%%cut here%%%%%%%
%$P_{\rm d} \approx 0.45/\rho^{1/2}$ (Frank, King \& Raine 1992).
%Assuming here $M_1 \approx 100 M_{\odot}$, we expect  
%characteristic periods $P \approx 5$--$7$ d. 
%Newton, I. 1687, Philosophiae Naturalis Principia Mathematica, 
%(London: S. Pepys), Vol 3, 419
%The System Of The World
%that the companion star intercepts $\sim 3$--$5\%$ of the X-ray luminosity, 
%this would result in an intercepted X-ray luminosity 
%of $\sim 2$--$3 \times 10^{39}$ erg s$^{-1}$ (Paper I), 
%to be partly reflected and partly thermalised 
%and re-radiated in the optical/UV bands.
%Even for a more massive primary, $M_1 \approx 500 M_{\odot}$, 
%the intercepted radiation would still 
%be $\approx 10^{39}$ erg s$^{-1}$.
%%%%%%%%%%%cut here%%%%%%%
The intercepted flux depends of course on the geometry of emission: 
it will be lower than the values estimated above 
if the X-ray emission is beamed in 
the direction normal to the orbital plane. However, 
it is possible that the intercepted flux is at least 
of the same order of magnitude as the intrinsic emission from 
the star ($L_{\rm bol} \approx 2$--$5 \times 10^{38}$ erg s$^{-1}$ 
in the BSG phase), causing the star to appear bluer and brighter 
than an isolated star of the same mass and age.
There is of course no firm evidence yet for this scenario, 
so ours is only a speculation, at this stage.

\begin{figure}
%  \vspace*{174pt}
\epsfig{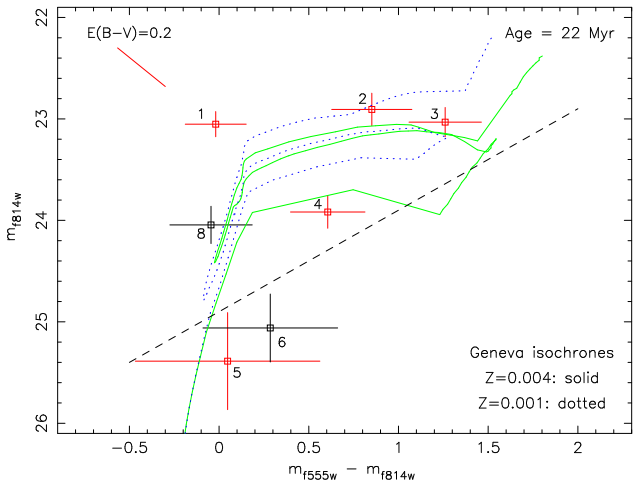}
  \caption{Colour-magnitude diagram and Geneva isochrones 
for seven candidate optical counterparts of the X7 ULX. 
An age $\approx 22$ Myr is consistent (within $1 \sigma$) 
with six out of seven, for $Z = 0.004$. 
The initial-mass spread along this isochrone 
is from $\approx 10.5 M_{\odot}$ (near the location 
of star No.~5) 
to $\approx 11.1 M_{\odot}$ (near the location of 
the RSG stars No.~2, 3).}
\end{figure}

Another piece of evidence in favour of star No.~1 as the true 
optical counterpart comes from a 1200s continuum-subtracted 
H$\alpha$ image (Fig.~13) taken by M. Pakull \& L. Mirioni 
with the 3.6m Canada France Hawaii Telescope (CHFT), 
on 2000 March 11. Although it has lower spatial resolution 
than the {\it HST}/WFPC2 images, it clearly reveals line emission 
associated with star No.~1. The image was not flux calibrated, 
hence we cannot measure the emitted H$\alpha$ flux precisely. 
However, we shall give an order-of-magnitude estimate in Sect. 5.2.
Detection of H$\alpha$ emission alone does not of course prove 
that star No.~1 is the ULX counterpart. We shall discuss 
possible scenarios for Balmer line emission in Sect. 5.2, 
some of them related to accretion or outflows near 
the compact object in an X-ray binary, some unrelated 
(e.g., colliding winds, winds from O-type supergiants, 
disks around Oe/Be stars).

%Variations in the H$\alpha$ emission associated 
%with X-ray state transitions would provide a significant 
%test. We are currently planning further photometric 
%and spectroscopic studies of this star.

Finally, we searched for possible optical variabilities 
in each of the five candidate stars.
Each {\it HST}/WFPC2 observation was in fact a series of four consecutive 
500-s exposures: we looked at the individual frames, but 
found no variability for any of the possible optical 
counterparts. Ellipsoidal modulations with an amplitude 
$\approx 0.2$ mag are expected from such systems 
(in analogy with, for example, the B-type optical counterpart 
of LMC X-3: Kuiper, van Paradijs \& van der Klis 1988); 
the amplitude could be higher 
if X-ray irradiation effects on the surface of the companion star 
are also important. Detection of such modulations 
is well within the capabilities of the WFPC2 or 
the Advanced Camera for Surveys (ACS) on board {\it HST}. 
A series of exposures over a longer period 
of time (a few days) is needed to investigate this possibility, 
which would strongly constrain the binary period.
%Kuiper, L.; van Paradijs, J.; van der Klis, M.1988A&A...203...79K

\subsection{Mass transfer in the ULX}

The observed X-ray luminosity of NGC\,4559 X7 (Paper I) 
implies an accretion rate onto the BH of 
$\approx 3.5$--$5 \times 10^{-6} M_\odot$~yr$^{-1}$, 
assuming isotropic emission and an efficiency $\approx 0.1$.
If the bolometric luminosity is considered, 
an accretion rate $\ga 10^{-5} M_\odot$~yr$^{-1}$ 
is required.

Massive supergiant stars are in an evolutionary 
phase at which a strong stellar wind drives 
a large mass loss.
If star No.~1 has a mass $\approx 20$--$25 M_{\odot}$, 
its mass loss rate in the stellar wind is 
$\approx 1$--$2 \times 10^{-6}M_\odot$~yr$^{-1}$~
($\sim 10^{20}$~g~s$^{-1}$) in the BSG stage 
(Lejeune \& Schaerer 2001). 
This rate is sustained for a timescale $\approx 5 \times 10^5$~yr.
As the star evolves towards the RSG stage, the mass loss 
rate in the wind will eventually increase 
to $\approx 10^{-5} M_\odot$~yr$^{-1}$ 
for a timescale of $\approx 10^5$~yr.
If the donor star is close to filling its Roche lobe, 
  the stellar wind is focused, 
  and as a consequence almost all the wind material will be accreted 
  onto the BH; this situation occurs for example 
in the Galactic source Cyg X-1 
  (Tarasov, Brocksopp \& Lyuty 2003). 
%  and it is the more extreme version.  
Such rates of mass transfer in a focussed wind 
can produce X-rays luminosities typical of moderately bright 
ULXs. However, the mass loss rate expected from star No.~1 
is not sufficient to explain the isotropic luminosity 
of NGC\,4559 X7.  
Moreover, a stellar wind is certainly not strong enough 
to account for the observed luminosity if the donor star 
in the X7 system has a mass $\approx 10$--$15 M_{\odot}$.

A higher accretion rate can be obtained 
  if the donor is directly overflowing its Roche lobe  
(e.g., Wellstein, Langer \& Braun 2001; 
Vanbeveren, De Loore \& Van Rensberg 1998).
In this case, all the mass lost from the donor star 
  will be captured by the BH. 
The BH candidate LMC X-3 is an example 
of an X-ray binary containing 
an early-type star transferring mass onto 
a (slightly) more massive BH via Roche lobe overflow 
(e.g., Wu et al. 2001).
For a $\sim 20 M_{\odot}$ star, peak mass transfer rates 
of $\sim 10^{-4} M_\odot$~yr$^{-1}$ 
can be achieved, over a thermal timescale $\sim 10^5$~yr.
(Podsiadlowski, Rappaport \& Han 2003; Wellstein et al.~2001). 
A proper determination of the accretion rate 
requires a study of the stability criteria 
for mass transfer as a function of mass ratio, 
evolutionary stage of the donor star, 
and rate of loss of angular momentum 
in a stellar or disk wind (Ritter 1988; Wu 1997).  
The stability and duration of mass transfer, in turns, 
affect the lifetime of the X-ray phase, 
  the mass transfer duty cycle, and the long- and short-term 
X-ray variabilities, and determine the relative abundance 
of active ULXs observable in young stellar environments 
at any given time.

We leave a quantitative investigation of the physical conditions 
for stable Roche-lobe mass transfer in ULXs, 
for a range of possible mass ratios, to a follow-up paper.
Here we simply point out one phenomenon 
that may be significant for our understanding of 
mass transfer in ULXs. In a binary system where mass 
is transferred from the less massive to the more massive 
component (as we suppose being the case for most ULXs, 
and in particular for NGC\,4559 X7), 
the orbit widens, and therefore the Roche lobe of the donor 
star expands. If the donor is a main sequence or subgiant 
star, its adiabatic mass-radius exponent is $> 0$, i.e., 
its thermal equilibrium radius decreases 
as the star loses mass. Hence, the star will soon 
become detached from its Roche lobe and mass transfer will cease, 
until contact is regained. A loss of angular momentum 
from the system (for example through a wind) 
is required to counteract this effect 
and shrink the orbit, thus ensuring a steady mass transfer.
However, if the donor is a supergiant star, its 
mass-radius exponent is $< 0$, so its thermal equilibrium 
radius will increase as the star loses mass 
(Wellstein et al.~2001). Hence, a supergiant donor 
may remain in contact with its Roche lobe and keep transferring 
mass steadily onto a more massive BH at a high rate, 
throughout its lifetime (up to $\sim 10^6$ yr), 
even in the absence of angular momentum losses.
Thus, we speculate that an O- or B-type supergiant star 
transferring mass via Roche-lobe overflow onto a more 
massive BH could be a natural explanation for persistent ULXs. 
Incidentally, this argument also suggests that 
the optical counterpart of X7 is more likely to be 
one of the supergiant stars listed in Table 2, 
rather than any of the fainter main-sequence stars 
in the same field.

%Vanbeveren, D.; De Loore, C.; Van Rensbergen, W. 1998A&ARv...9...63V
%Tarasov, A. E.; Brocksopp, C.; Lyuty, V. M.2003A&A...402..237T
%Podsiadlowski, Ph.; Rappaport, S.; Han, Z. 2003MNRAS.341..385P

Finally, although we cannot yet identify the donor star, 
we can rule out an association of the ULX 
with a young, massive cluster. There is a small group 
of young stars but no bright clusters 
inside the X-ray error circle (Fig.~3): all the optical sources 
are consistent with single stars or, at most, 
the superposition of a few stars. The only moderately-large 
cluster in the field is located $\ga 570$ pc away 
(projected distance, without considering the viewing angle), 
on the south-eastern side of the star-forming complex.
Given the age of the cluster, a physical association 
would require an ejection speed $\ga 25$ km s$^{-1}$ for the ULX binary 
system. This value is too large to be consistent with 
an intermediate-mass BH (Zezas \& Fabbiano 2002).

%Zezas, A., \& Fabbiano, G. 2002, ApJ, 577, 726
%\vspace{1cm}

\section{Origin of the large star-forming complex}

\subsection{Mass, density and star-formation rate in the region}

The CHFT H$\alpha$ image shows more clearly 
the structure of the large star-forming complex 
around the ULX. The ring-like structure 
of the H{\footnotesize II} region (Figs.~13, 14) suggests 
that an expanding wave of star formation has recently moved 
from a centre (which appears to be near but not coincident 
with the ULX) outwards. 
Assuming a Salpeter IMF down to a lower mass limit 
of $1 M_{\odot}$, as explained in Sect. 4.3, 
we estimate from Starburst99 models (Leitherer et al. 1999) 
that continuous star formation over the last 30 Myr at a rate of 
$\sim 10^{-2}$--$10^{-2.5} M_{\odot}$ yr$^{-1}$ (depending 
on the choice of metal abundance and upper mass limit) 
would account for the integrated luminosity 
($M_B \approx -13.5$) and the observed number of 
O stars\footnote{If we assume a cut-off of the Salpeter IMF 
at $0.1 M_{\odot}$ instead, the inferred star formation rate is 
$\sim 10^{-1.5}$--$10^{-2} M_{\odot}$ yr$^{-1}$.}.

This rate implies a mass in stars 
of $\sim 1$--$3 \times 10^5 M_{\odot}$.
The efficiency for the conversion of molecular gas 
into stars is generally $\la 10\%$ (e.g., Tan 2000; 
Planesas, Colina \& Perez-Olea 1997). 
Moreover, some of the gas will be in atomic form; we have 
no observational data to determine the H{\footnotesize{I}}/H$_2$ ratio 
in this region, but we can assume for the sake of our 
order-of-magnitude calculation that they contribute 
in equal measure; we shall later see that this assumption 
is justified. Therefore, we estimate that the total mass 
(stars plus swept-up gas) in this complex is at least 
a few $10^6 M_{\odot}$.

On the other hand, we can also place an upper limit 
to the total mass in the region.
We see from the {\it XMM-Newton}/OM image (Fig.~1) that it is located 
just outside the star-forming region in the disk of NGC\,4559. 
This suggests that, before the triggering event, 
the local surface density was just below 
the threshold necessary for the onset of gravitational 
collapse and star formation. 
There are two empirical thresholds for star formation 
(Elmegreen 2002 and references therein).
A constant-density threshold (more often applied 
to dwarf and irregular galaxies) requires a total surface density 
$\Sigma_{\rm g} > \Sigma_{\rm min} \approx 6 M_{\odot}$ pc$^{-2}$.
Alternatively, a variable threshold based on Toomre's 
stability criterion (Toomre 1964) is more often applied 
to spiral galaxies:
\begin{displaymath}
\Sigma_{\rm g} > \Sigma_{\rm min} \approx \alpha \frac{\kappa \sigma}{\pi G},
\end{displaymath}
where $\alpha \approx 0.7$ is an empirical constant, 
$\sigma$ is the velocity dispersion of the gas, 
$\kappa = (2)^{0.5}\,(V/R)\,(1+dV/dR)^{0.5}$ is the 
epicyclic frequency at radial distance $R$, and 
$V$ is the circular velocity of the galaxy 
at radius $R$ (e.g., Kennicutt 1989, 1998; 
Wong \& Blitz 2002; Boissier et al.~2003). 
We can estimate $V(R)$ in NGC\,4559 from 
the WHISP Survey\footnote{http://www.astro.rug.nl/$\sim$whisp/} 
21-cm H{\footnotesize I} radio observations: at a distance 
$R = 16$ kpc, the velocity curve is flat, with an asymptotic velocity  
$V \approx 110 \pm 5$ km s$^{-1}$. The gas velocity 
dispersion $\approx 5$--$10$ km s$^{-1}$ (Boissier et al.~2003); 
a value commonly used is $\sigma = 6$ km s$^{-1}$, 
following Kennicutt (1989).
Hence, the Toomre criterion implies a local surface density 
$\Sigma_{\rm g} < \Sigma_{\rm min} \approx 4$--$7 M_{\odot}$ pc$^{-2}$ 
before the onset of star formation, 30 Myr ago; this is  
in agreement with the threshold estimated from the constant-density 
criterion.

The star-forming complex has a projected area on the sky 
of $\approx 300$ arcsec$^2$. Considering the viewing angle, 
this corresponds to a physical area of $\approx 1.8 \times 10^6$ pc$^2$.
From this, we can estimate that the total mass (essentially, 
atomic and molecular gas) contained 
in that region before the triggering of star formation 
was $< 10^7 M_{\odot}$. 
Combining this upper limit with the lower limit 
derived earlier, we can conclude that the star-forming 
complex contains a total mass $\approx 5$--$10 \times 10^6 M_{\odot}$.

We speculate that an initial perturbation 
caused a local increase in the gas density, 
triggering an expanding wave of star formation.
Considering the average star-formation rate 
estimated earlier, and the size of the ring 
where star formation has taken place, we can use 
a relation in Kennicutt (1998) (see, in particular, 
his Figure 6) to estimate the gas density in the ring: 
we infer that $\Sigma \approx 10$--$20 M_{\odot}$ pc$^{-2}$. 
Therefore, a small density perturbation, by a factor $\approx 2$, 
would have sufficed to trigger the star-forming event.

As an aside, we note that these values of the 
total gas surface density correspond to 
a pressure $\log (P/k_{\rm B}) \approx 3.5$--$4$ 
(where $P/k_{\rm B}$ is expressed in cm$^{-3}$ K) 
at the disk midplane, which in turns corresponds 
to a regime where molecular and atomic hydrogen give 
a similar contribution to the total gas mass (Wong \& Blitz 2002). 
This is consistent with our earlier assumption.

\subsection{H$\alpha$ luminosity}

Having estimated the total SFR of the complex, 
we can re-examine the CFHT H$\alpha$ image, recalling the 
relation between SFR and H$\alpha$ luminosity 
in star-forming galaxies, namely 
$L_{{\rm H}\alpha}\,({\rm erg~s}^{-1}) \approx 10^{41.3 \pm 0.2}$ SFR 
($M_{\odot}$~yr$^{-1}$) (Buat et al.~2002; Kennicutt 1998).
We expect an H$\alpha$ luminosity $\approx 10^{39}$ erg s$^{-1}$ 
for the whole complex, neglecting the effect 
of extinction for this order-of-magnitude calculation. 

Alternatively, from our previous estimate of the total 
mass in the star-forming complex, and the size 
of the H$\alpha$-emitting ring, we estimate 
a baryon number density $\sim 1$ cm$^{-3}$ 
in that ring. This estimate can only be accurate to an order 
of magnitude, since we ignore the real geometry 
and vertical extent of that region.
Assuming that H$\alpha$ comes from Case B recombinations 
in a gas at $T \approx 10,000$ K (typical of warm interstellar 
medium), we can determine the emission coefficient $j_{{\rm H}\alpha}$, 
and the luminosity $L_{{\rm H}\alpha} \approx 4\pi\,j_{{\rm H}\alpha}$.
We have $4\pi\,j_{{\rm H}\alpha}/(N_p N_e) \approx 3 \times 10^{-25}$ 
erg s$^{-1}$ cm$^{3}$ (Osterbrock 1989). This value 
is only weakly dependent on the temperature, changing 
by a factor of 3 over the $5,000$--$20,000$ K temperature range.
Integrated over a characteristic scale $\sim (500 \ \rm{pc})^3$, 
this implies a total H$\alpha$ luminosity $\sim 10^{39}$ erg s$^{-1}$, 
consistent with our previous estimate.

Using this value as a calibration for the total observed 
count rate in the image, we can then 
estimate an H$\alpha$ luminosity $\approx 1$--$3 \times 10^{35}$ 
erg s$^{-1}$ for the optical counterpart of the ULX.
We stress that this is only an order-of-magnitude estimate, 
given the large number of assumptions on which it is based.
We will carry out a more accurate study of the optical 
continuum and H$\alpha$ emission with the {\it HST}/ACS, next year.

As discussed earlier, a small group of O and early-type B stars 
is located at the ULX position. Are they the Balmer line sources 
seen in the CHFT image?
H$\alpha$ emission is often detected from the excretion 
disk or envelope around Oe/Be stars: the brightest 
systems have a luminosity $\approx 2$--$4 \times 10^{34}$ 
erg s$^{-1}$ (Apparao 1998; Stevens, Coe \& Buckley~1999; 
Janot-Pacheco, Motch \& Pakull 1988), 
still a few times fainter than observed at the X-7 position 
in NGC\,4559. OB supergiants are another source 
of Balmer emission: the H$\alpha$ line emission is proportional 
to the mass loss rate in the stellar wind, 
and can reach luminosities $\approx 7 \times 10^{34}$ 
erg s$^{-1}$ for stars with a mass-loss rate of $\approx 10^{-5.4} 
M_{\odot}$ yr$^{-1}$ (Klein \& Castor 1978; Ebbets 1982).
%Ebbets, D. 1982ApJS...48..399E
%	Klein, R. I.; Castor, J. I. 1978ApJ...220..902K
Colliding winds in O-type close binary systems are also 
a source of Balmer emission: typical luminosities are 
$\sim$ a few $10^{34}$ erg s$^{-1}$ (Thaller et al.~2001). 
%Thaller, M. L.; Gies, D. R.; Fullerton, A. W.; Kaper, L.; Wiemker, R.
%2001ApJ...554.1070T
We conclude that these mechanisms are probably too faint 
to explained the observed H$\alpha$ flux at the ULX position.

None of the previous scenarios requires the presence of an accreting 
compact object. However, a compact object may provide 
additional mechanisms for H$\alpha$ emission.
Line emission is usually detected from accretion disks 
in low-mass X-ray binaries; however, characteristic luminosities 
are generally much lower: for example, $L_{{\rm H}\alpha} \sim$ a few 
$\times 10^{31}$ erg s$^{-1}$ for the BH candidate 
GRO 1955$-$40 during the June 1996 high-soft state 
(Soria, Wu \& Hunstead 2000).
Stronger emission is seen from the high-mass X-ray binary 
Cyg X-1, where $L_{{\rm H}\alpha} \approx 2 \times 10^{33}$ 
erg s$^{-1}$ in the high/soft state (Tarasov et al. 2003).
%Tarasov, A. E.; Brocksopp, C.; Lyuty, V. M. 2003A&A...402..237T
For higher values of the mass transfer rate, close 
or above the Eddington accretion rate, a strong radiatively-driven 
wind is likely to be formed from the accretion disk surface, 
resulting in an optically-thick envelope 
or outflow enshrouding the X-ray source (Shakura \& Sunyaev 1973).
For example, the intermediate-mass X-ray binary V4641 Sgr 
shows strong H$\alpha$ emission ($L_{{\rm H}\alpha} \approx 4 
\times 10^{34}$ erg s$^{-1}$ during a short phase 
of super-Eddington accretion (Revnivtsev et al 2002a,b; Chaty et al.~2003).
This is still less than that observed 
near the ULX position.

Even stronger Balmer line emission is found in the 
BH candidate SS\,433, a system for which 
persistent super-Eddington accretion has been invoked: 
the luminosity in the ``stationary'' H$\alpha$ component 
(thought to be mostly associated with the inner accretion disk 
and its envelope) is $L_{{\rm H}\alpha} \sim 10^{36}$ erg s$^{-1}$ 
(Asadullaev et al.~1983). 
Super-Eddington accretion onto a stellar-mass BH, 
with the formation of strong radiatively-driven disk winds 
and outflows, is a possible explanation for ULXs 
with X-ray luminosities up to $\approx 10^{40}$ erg s$^{-1}$ 
observed in star-forming environments (King 2002, 2003).
It was also suggested that the soft thermal component 
at $kT \sim 0.1$ keV often seen in those systems 
(including NGC\,4559 X-7: Paper I) 
may come from the down-scattering of harder X-ray photons 
in the outflow (King \& Pounds 2003).

Another possibility is that the H$\alpha$ emission 
comes from X-ray ionized clouds around the ULX, 
a scenario similar to what is found in Seyfert galaxies.
We shall estimate the order-of-magnitude H$\alpha$ luminosity 
expected from X-ray ionized gas in the next section.
In any case, future studies of the 
H$\alpha$ emission from the X7 counterpart 
may reveal more clues on the nature of this X-ray source.
In particular, variations in the H$\alpha$ emission associated 
with X-ray state transitions would provide a significant test 
for the size, geometry and true luminosity of the system.

\subsection{An X-ray ionised nebula?}

In addition to a possible point-like counterpart, 
we should also consider whether the ULX contributes 
to the ionisation of the larger ring-like structure.
Emission nebulae with sizes of a few hundred pc, 
showing both low and high ionisation lines, with 
$L_{{\rm H}\alpha} \sim$ a few $\times 10^{37}$--$10^{38}$ erg s$^{-1}$, 
have been found around many nearby ULXs, for example 
NGC\,1313 X-1 and X-2, M\,81 X-9, Holmberg II X-1 
(Pakull \& Mirioni 2002; Wang 2002). Some of these nebulae
have been shown to be X-ray photoionised by the ULX. 
On a smaller scale (size of a few pc, $L_{{\rm H}\alpha} 
\approx 10^{37}$ erg s$^{-1}$), an X-ray photoionised 
emission nebula was found around the BH candidate 
LMC X-1 (Pakull \& Angebault 1986).

Extrapolating the X-ray spectral fits presented in Paper I, 
we can estimate the number of photons of energy $> 13.6$ eV 
intercepted by the gas surrounding the X-ray source 
(i.e., not including the line-of-sight foreground absorption), 
assuming that the emission is isotropic.
Blackbody-plus-powerlaw models give the highest number 
of ionising photons ($\approx 3 \times 10^{49}$ s$^{-1}$); 
however, this is almost certainly an upper limit, because 
the power-law component is likely to be truncated at low energies. 
Comptonisation models such as {\tt bmc} may provide 
a more physical estimate: in this class of models, 
the power-law component does not extend below the seed 
thermal component at $kT \approx 0.1$ keV. Using the 
{\tt bmc} fit to the {\it XMM-Newton} spectrum (Paper I), 
we estimate $\approx 10^{49}$ ionising photons s$^{-1}$ 
from the ULX, intercepted by the surrounding nebula 
along our line of sight. The absorption in the NGC\,4559 
galactic plane is likely to be higher; if all the X-ray flux 
were intercepted by the nebula there, it would 
approximately double the supply of Hydrogen-ionising 
photons, i.e., $\approx 2 \times 10^{49}$ photons s$^{-1}$ 
using the {\tt bmc} spectral model.

The number of H$\alpha$ photons emitted as a consequence 
of this ionising flux is given by:
\begin{eqnarray*}
\frac{L_{{\rm H}\alpha}}{hv_{{\rm H}\alpha}} & \approx &
\frac{\alpha^{{\rm{eff}}}_{{\rm H}\alpha}(H^0,\,T)}{\alpha_{{\rm B}}(H^0,\,T)}\,
\int^{\infty}_{13.6 \, \rm{eV}}\frac{L_{\epsilon}}{\epsilon}\,
[1-\exp(-\tau_{\epsilon})]\,{\rm d}\epsilon \\
& \approx & \frac{\alpha^{\rm eff}_{{\rm H}\alpha}}{\alpha_{{\rm B}}} \times 
(1-3) \, 10^{49} \ \rm{s}^{-1},
\end{eqnarray*}
where $\tau_{\epsilon}$ is the total optical depth in the nebula 
at energy $\epsilon$, 
$\alpha^{{\rm{eff}}}_{{\rm H}\alpha}(H^0,\,T)$ is the 
effective recombination coefficient for the H$\alpha$ line, 
and $\alpha_{{\rm B}}(H^0,\,T)$ is the recombination coefficient 
summed over all levels above the ground state (Pakull \& Angebault 1986, 
and references therein). The ratio of the recombination 
coefficients is only weakly dependent on the gas temperature; using 
typical values from Osterbrock (1989), we obtain  
an H$\alpha$ line emission of $\approx (3$--$10) \times 10^{48}$ 
photons s$^{-1}$, i.e., a luminosity $\approx (1$--$3) \times 10^{37}$ 
erg s$^{-1}$. This is much less than the total H$\alpha$ luminosity 
estimated earlier, for the whole star-forming ring. 
We conclude that the X-ray flux from the ULX does not 
significantly contribute to the Balmer-line emission 
of the large surrounding H{\footnotesize{II}} region.
However, it may explain the point-like emission 
seen within the ULX error circle, as discussed 
in the previous subsection.
 
Detection of He{\footnotesize{II}}\,$\lambda 4686$ line emission 
would provide a much more stringent test for the presence of 
X-ray photoionised gas around the ULX: this is because 
even the hottest O stars do not provide enough 
photons with energies $> 54$ eV, required to produce 
doubly-ionised Helium. Measurements of high 
[O{\footnotesize{I}}]\,$\lambda 6300$/H$\alpha$ 
and [O{\footnotesize{III}}]\,$\lambda 5007$/H$\beta$ 
emission line ratios would be another test for 
the X-ray photoionisation scenario  
(Pakull \& Mirioni 2002). We are planning to conduct 
new spectroscopic observations of the field, 
to investigate this issue.

%We also note, as an aside, that the ring-like structure 
%is more evident in the H$\alpha$ emission than 
%in the stellar distribution. 

%We are currently planning further photometric 
%and spectroscopic studies of this star.

%Buat, V.; Boselli, A.; Gavazzi, G.; Bonfanti, C 2002A&A...383..801B

%Asadullaev, S. S.; Aslanov, A. A.; Kornilov, V. G.; Cherepashchuk, A. M.
%1983SvAL....9..282A
%This envelope is likely the result of a near- or super-Eddington 
%rate of mass accretion onto the black hole. The envelope vanishes 
%during subsequent evolution of the source when the apparent luminosity 
%drops well below the Eddington value. Thus this transient source p
%rovides us with direct proof of the dramatic change in the character 
%of an accretion flow at the mass accretion rate near or above 
%the critical Eddington value as predicted long ago by the theoretical models.

\subsection{What triggered the expanding wave of star formation?}

If we interpret the complex as an expanding star-formation front, 
an important question to address is what the relation 
is between the ULX and the surrounding H{\footnotesize{II}} 
region. Large, isolated shell- or ring-like 
star-forming complexes of comparable size ($500$--$1000$ pc) 
and age ($10$--$30$ Myr) have been found in other nearby 
spiral galaxies: for example in NGC\,6946 (Larsen et al. 2002), 
and, on a smaller scale, in M\,83 (Comer\'{o}n 2001). Gould's Belt 
in the Milky Way is also similar.
The star-forming complex in NGC\,4559 is  
a factor of two brighter than Gould's Belt and a factor of four fainter 
than the NGC\,6946 complex. The latter has a young super-star 
cluster at its centre, while Gould's Belt and 
the complex in M\,83 contain only OB associations. 
None of them contains a ULX, unlike the star-forming complex 
in NGC\,4559. This suggests that the ULX 
is a side effect, rather than the determinant factor 
for the triggering and subsequent evolution 
of the star-forming episode.

Possible explanations for the initial triggering of such 
complexes are (Elmegreen, Efremov \& Larsen 2000; 
Larsen et al.~2002): the collapse of a "supergiant molecular cloud" 
at the end of a spiral arm; a hypernova explosion (which  
in our case might also have been the progenitor of NGC\,4559 X7); 
or the infall of a high-velocity H{\footnotesize I} 
cloud or satellite galaxy through 
the outer galactic disk. In all cases, the initial perturbation 
creates a radially expanding density wave or ionization front, 
which sweeps up the neutral interstellar medium. Clustered 
star formation along the expanding bubble rim is triggered by 
the gravitational collapse of the swept-up material 
(e.g., Elmegreen \& Lada 1977; Whitworth et al.~1994).

The large size of the complex in NGC\,4559, its location 
in the outer disk, the lack of other star-forming regions 
nearby, and the absence of diffuse 
X-ray emitting gas inside the star-forming complex seem 
to favour the collision hypothesis over the hypernova model 
(Tenorio-Tagle et al.~1986, 1987).
%In particular, the X7 complex is located just beyond 
%the outer star-forming radius in the NGC\,4559 disk (Fig.~1):  
%we expect that region to be only marginally stable 
%against spontaneous gravitational collapse 
%($Q \ga 1$: Binney \& Tremaine 1987; Toomre 1963). 
%Hence, a small density perturbation may suffice to induce 
%local gravitational collapse of the interstellar gas.
The impact of an $\approx 3 \times 10^5 M_{\odot}$ H{\footnotesize I} cloud 
on the Milky Way disk was simulated (Comer\'{o}n \& Torra 1994) 
to explain the formation of Gould's Belt. They show 
that, after $\approx 30$ Myr, the mass of the cold swept-up 
material is comparable or larger than the mass of the impacting cloud.
Similar consequences would come from the passage 
of a globular cluster ($M \sim 10^6 M_{\odot}$) 
through a gas-rich disk (Wallin, Higdon \& Staveley-Smith 1996).
In that case, the perturbation imparted by the colliding object 
to the disk gas would be entirely gravitational.
A proper motion study of the Milky Way globular cluster NGC\,6397 
(Rees \& Cudworth 2003) found that the cluster passed through 
the Galactic disk $\la 5$ Myr ago, and may have dynamically 
triggered the formation of the young open cluster NGC\,6231. 

%Rees, R. F.; Cudworth, K. M. American Astronomical Society Meeting 203, #10.06

%Begum, Ayesha; Chengalur, Jayaram N.; Hopp, Ulrich2003NewA....8..267B
%Camelopardalis B

\begin{figure}
%  \vspace*{174pt}
\psfig{figure=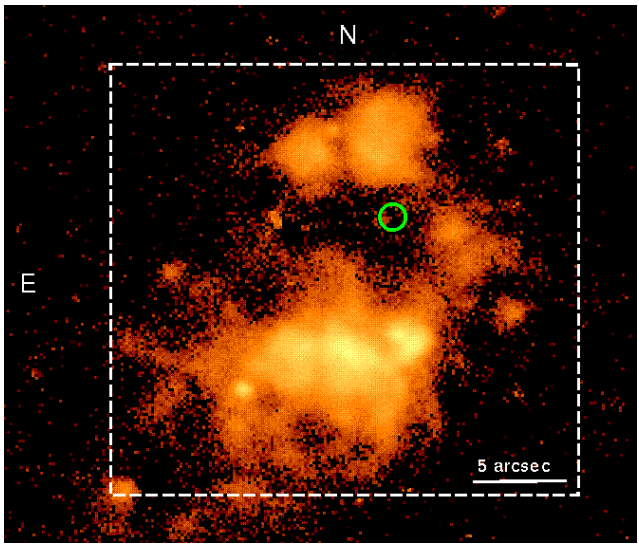,width=84mm}
  \caption{H$\alpha$ image from the 3.6-m CHFT. The H$\alpha$ 
contour lines for the region inside the dashed box 
are shown in Fig.~14.}
\end{figure}

\begin{figure}
%  \vspace*{174pt}
\psfig{figure=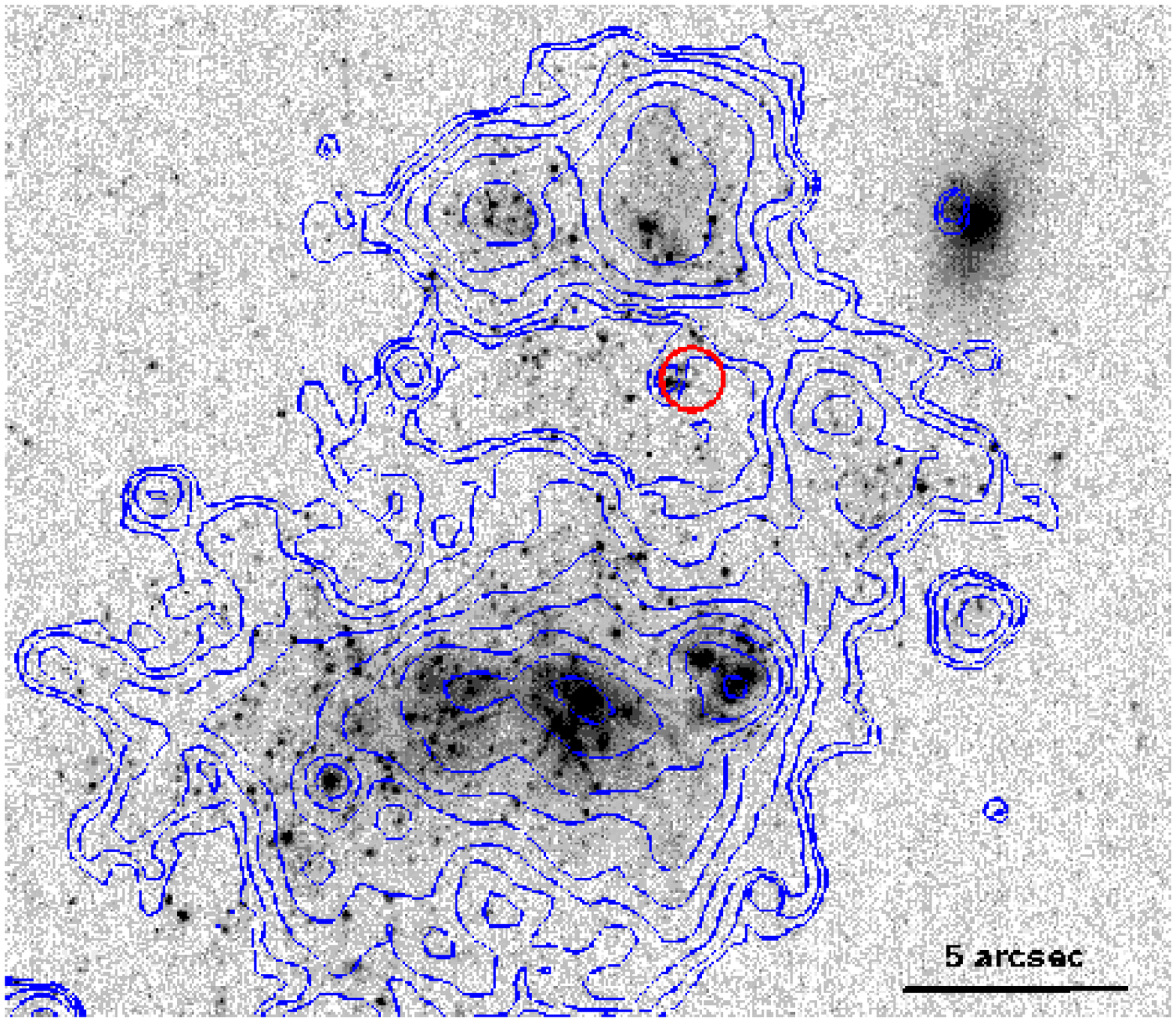,width=84mm}
\psfig{figure=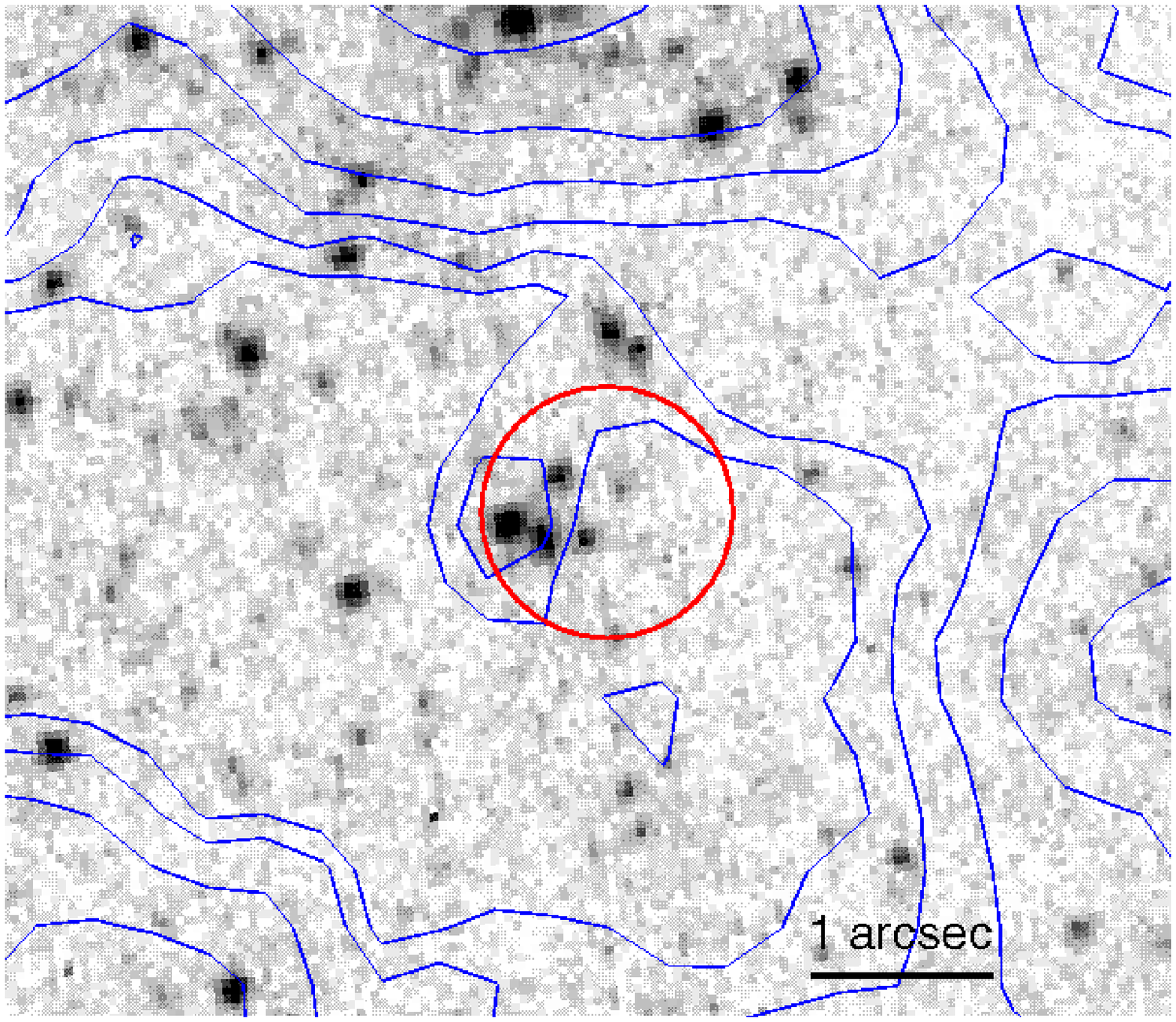,width=84mm}
  \caption{The ULX is located inside, although not at the center, 
of a large ring-like HII region. Greyscale image: {\it HST}/WFPC2 (PC chip), 
f450w filter ($\approx B$). Contours: H$\alpha$ image from the 
3.6-m CHFT. We speculate that the the dIrr galaxy seen 
at the north-west corner 
of the field is responsible for triggering this expanding 
wave of star formation, as it plunged through the gas-rich 
outer disk of NGC\,4559 some 30 Myr ago.}
\end{figure}

\subsection{A dwarf galaxy plunging through the disk?}
\label{sec:cartwheel}

An intriguing result of our optical study 
of the X7 environment is that we do indeed see an object 
that could have plunged through the gas-rich disk of NGC\,4559, 
triggering the expanding density wave.
The culprit could be the yellow galaxy located $\approx 7$\arcsec\  
(projected distance of $340$ pc) north-west of the ULX (Figs.~2, 14, 15).
This object cannot be a large background elliptical galaxy, because 
its isophotes are too irregular (Fig.~15). Its shape, size and luminosity 
are consistent with a dwarf irregular (or, possibly, a tidally-disturbed 
dwarf elliptical) at approximately the same distance 
as NGC\,4559 (Sandage \& Binggeli 1984). 
We cannot presently rule out the possibility 
that it is a chance line-of-sight coincidence, but 
we suggest that the most natural interpretation  
is a small satellite galaxy of NGC\,4559.
Existing 21cm H{\footnotesize I} radio observations 
(e.g., the WHISP survey) 
do not show any large-scale velocity distortions, 
but the satellite dwarf is perhaps too small to influence 
the global galactic kinematics significantly.
However, as it crossed the disk, it would have 
significantly perturbed the local gas density, both 
hydrodynamically (ram pressure) and dynamically 
(gravitational perturbation). A dIrr galaxy contains 
both gas and dark matter, hence its impact would have been 
even more significant than that of a globular cluster. 
We are planning optical spectroscopic observations 
to determine the kinematics and distance 
of the dwarf galaxy, and therefore test this hypothesis.

If the dIrr galaxy is indeed physically associated 
with the star-forming complex, its integrated absolute brightness 
is $M_B \approx -10.7$, with observed color indices 
$B-V \approx 0.47$ and $V-I \approx 0.73$.
These colors are typical of a population dominated 
by F5--F8 main-sequence stars, suggesting an old age. 
Assuming a single burst of star formation, 
we infer (using Starburst99, Leitherer et al.~1999) 
a visible stellar mass of $\sim 10^6 M_{\odot}$ 
for the galaxy and an age $\ga 10^9$ yr for the dominant 
component of its stellar population.

As a comparison, this candidate satellite galaxy 
would be somewhat similar to the faint dwarf irregular 
galaxy Camelopardalis B (Cam B) in the IC\,342/Maffei group, 
which has $M_B \approx -10.9$, 
$B-V \approx 0.6$ (Begum, Chengalur \& Hopp 2003). 
The stellar mass in Cam B is estimated to be 
$\approx 3.5 \times 10^6 M_{\odot}$, 
and the total gas mass $\approx 6.6 \times 10^6 M_{\odot}$;
the total dynamical mass is $\approx 1.1 \times 10^8 M_{\odot}$, 
hence the galaxy is strongly dark-matter dominated 
(Begum et al.~2003). In fact, dark matter density 
dominates over stellar and gas density even in the inner regions.

\subsection{Star formation induced by the colliding satellite?}
\label{sec:cartwheel}

In addition to the old stellar component, 
the candidate satellite galaxy near NGC\,4559 shows 
two bright clusters and a few more, much fainter lumps.
The two brightest clusters have luminosities $M_B \approx -7.2$ 
and $M_B \approx -7.1$, and colors consistent 
with an age $\sim 10^7$ yr and masses of $\sim$ a few $\times 
10^3 M_{\odot}$. Their brightness and morphology 
is consistent with the star-forming complexes often found 
in dIrr galaxies (Parodi \& Binggeli 2003).
We obtain that the percentage of flux in the $B$ band 
due to these star-forming complexes (``lumpiness index'') 
is $\approx 7\%$ of the total B-band flux, 
the same value found for a large 
sample of irregulars and spirals regardless of Hubble type 
and galactic mass (Elmegreen \& Salzer 1999; Parodi \& Binggeli 2003).
Thus, the lumpiness index is thought to be a measure 
of star-formation efficiency. These considerations 
support the idea that the bright lumps are indeed 
clusters in the dIrr satellite galaxy and not 
simply background or foreground stars in NGC\,4559.
The dwarf galaxy is also detected in the {\it XMM-Newton}/OM image 
(Fig.~16), taken in the UVW1 filter ($\approx 2500$--$3500$ \AA).
This also indicates recent star formation, in addition 
to the old stellar population which would not be detected 
in the near-UV. (On the other hand, the H$\alpha$ image, 
Fig.~13, shows that the current star formation rate 
is negligible). 
The OM count rate is $0.115\pm0.005$ ct s$^{-1}$: if we assume 
that the colors of the younger population are typical of a B0V star, 
such a count rate in the UVW1 filter corresponds to 
$M_B \approx -7.9\pm0.1$, in agreement with the {\it HST}/WFPC2 
observations.

It is possible that this small, later episode of star formation 
in the dIrr galaxy may also have been triggered when this small 
satellite passed through the disk of NGC\,4559. 
If so, the younger stars may have been formed either 
from gas left in the dwarf galaxy, or collected 
into its gravitational potential well as it 
passed through the gas-rich disk.
%shocking its gas and creating an expanding star-forming wave. 
The non-spherical appearance 
of the large star-forming complex in NGC\,4559 may be due to 
an oblique impact, possibly from the south-east to the 
north-west direction. 
%(because the oldest 
%stars in the field are found in the south-east 
%sector, with ages $\ga 20$ Myr; Sect.~3.1).
%The relative masses of the ``bullet'' and swept-up gas 
%are consistent with the results of Comer\'{o}n \& Torra (1994) 
%for the case of Gould's Belt. 
The projected distance of the dIrr from the centre 
of the star-forming ring is $\approx 400$ pc, corresponding 
to a projected relative velocity of $\approx 15$ km s$^{-1}$ 
over 30 Myr; however, this is certainly a lower limit, 
since we have no elements to estimate the 
velocity component along the line of sight.

Thus, we could view the star-forming complex in NGC\,4559 
as a small-scale version of the Cartwheel galaxy, 
where many young ULXs have been detected in the expanding, 
star-forming ring (Gao et al.~2003).
%Apart from the different time and length scales involved, 
%the main difference between the two systems is that, 
In the Cartwheel, the initial perturbation causing 
the expanding density wave is entirely due to the gravitational 
interaction between the two galaxies; in the case of NGC\,4559, it 
may be due both to the dynamical impulse and to 
the direct hydrodynamical interaction between 
the gas in the satellite and in the disk.
A more detailed analysis of this possibility is left 
to a separate work.

\begin{figure}
%  \vspace*{174pt}
\psfig{figure=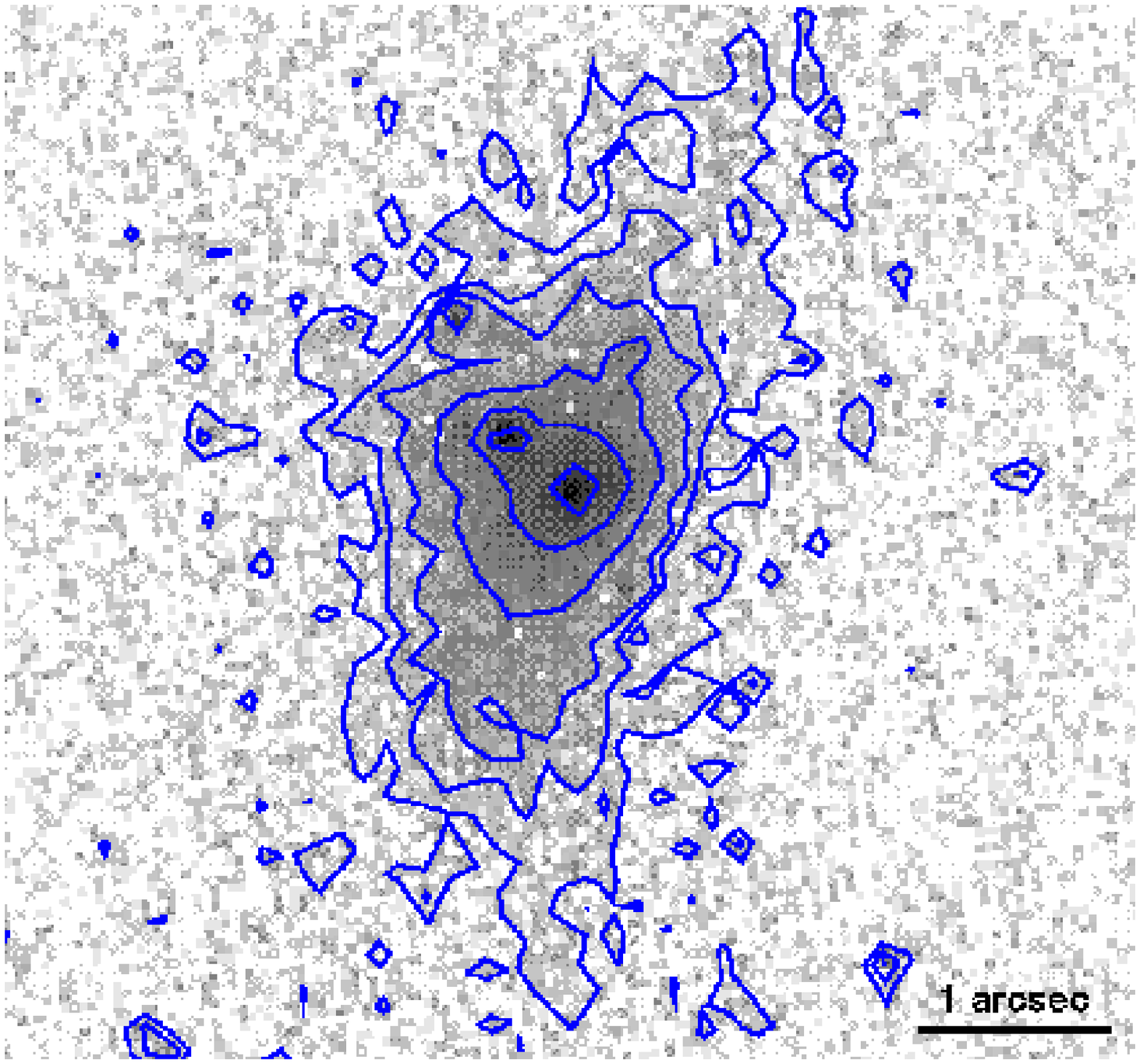,width=84mm}
\psfig{figure=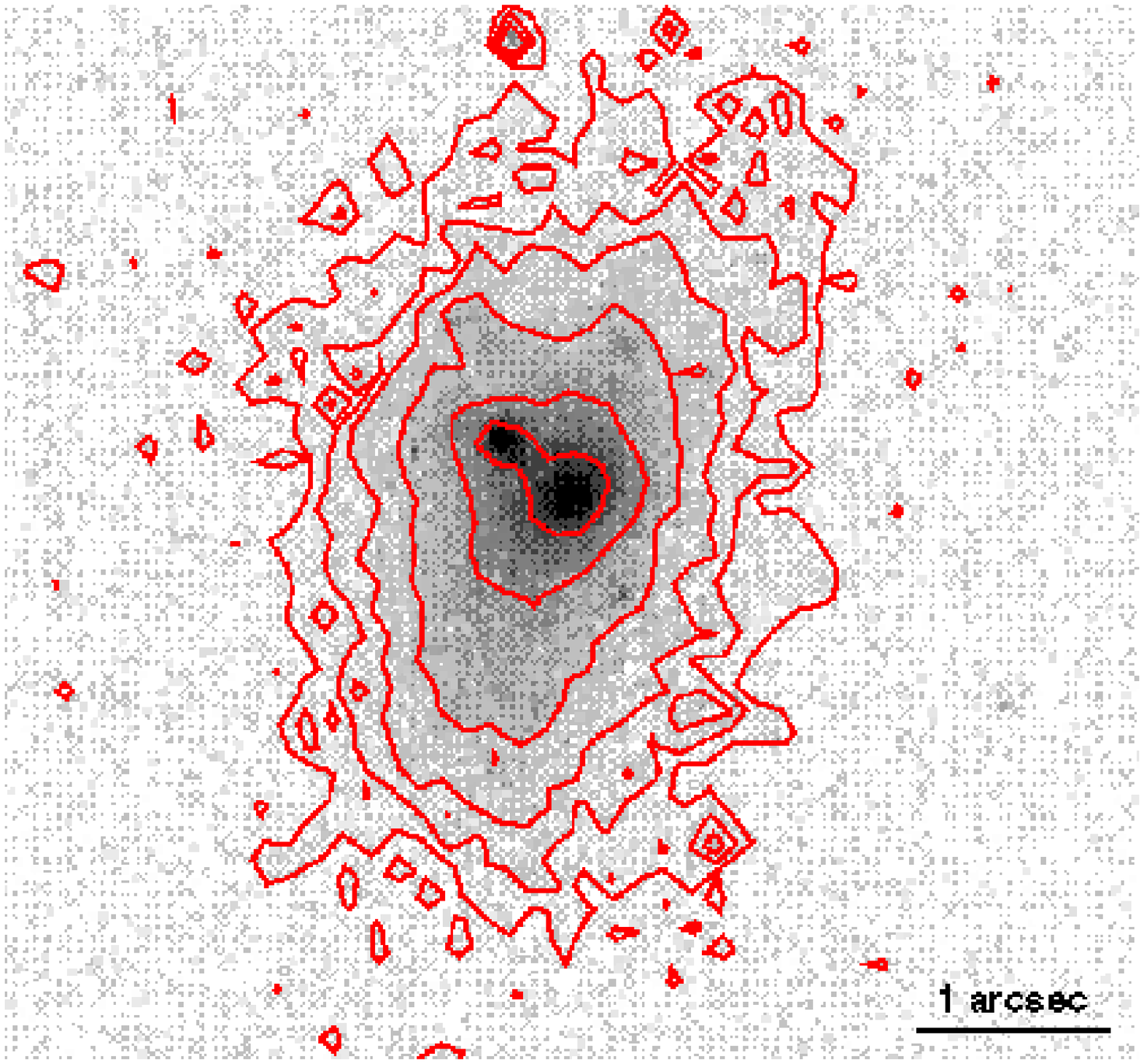,width=84mm}
  \caption{Close-up images of the candidate satellite galaxy, 
from the {\it HST}/WFPC2 images. 
Top panel: image in the f450w filter; bottom panel: in the f814w filter.
Both sets of contours are in a logarithmic scale, with arbitrary zeropoint.}
\end{figure}

\begin{figure}
%  \vspace*{174pt}
\psfig{figure=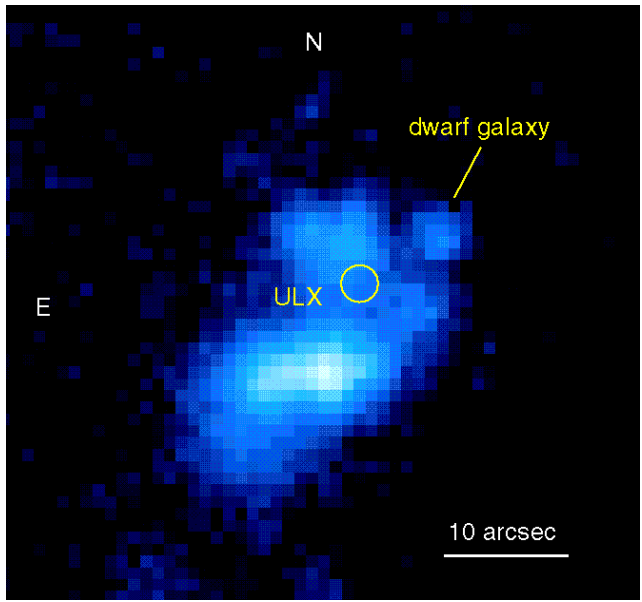,width=84mm}
  \caption{The dwarf galaxy near the X7 ULX is detected in the 
{\it XMM-Newton}/OM image (UVW1 filter). This is evidence 
of recent (but not current: cf. the H$\alpha$ image, Figs.~13, 14), 
moderate star formation.}
\end{figure}

\section{Environmental conditions favourable to ULX formation}
\label{sec:environs}

NGC\,4559 X7 offers an example of a bright ($L_{\rm x} > 2 \times 
10^{40}$ erg s$^{-1}$) 
ULX in a low-metallicity, actively star-forming environment 
disturbed by close galaxy interactions. 
At least one, and often all of these elements 
seem to be a common feature for many of the galaxies 
hosting ULXs: for example, galactic interactions 
and high star-formation rates 
for the Antennae (Zezas et al.~2002; Fabbiano et al.~2003), 
the Cartwheel ring (Gao et al.~2003), M\,82 (Griffiths et al.~2000), 
the Mice (Read 2003), NGC\,7714/15 (Soria \& Motch 2004); 
low metal abundance for the Cartwheel ring, NGC\,7714/15, 
the M\,81 group dwarfs (Wang 2002; 
Makarova et al.~2002), and a group of nearby 
blue compact dwarfs including I\,Zw\,18 (Thuan et al.~2004). 
A preferential association between ULXs and interacting 
galaxies was shown in Swartz et al.~(2004); a connection 
between ULX formation and low-Z environments was suggested 
in Pakull \& Mirioni (2002). 

%Assuming that most ULXs can be explained by accreting 
%black holes more massive ($\approx 50$--$100 M_{\odot}$) 
%than those found in nearby X-ray binaries, 
This leads us to speculate that these two environmental conditions 
may be particularly favourable for producing massive remnants.
For example, galaxy mergers, close interactions, 
and collisions with satellite galaxies or high-velocity 
H{\footnotesize I} clouds favour clustered star formation. 
The core of young star clusters is an environment 
where intermediate-mass BHs (with masses up 
to $\sim 1,000 M_{\odot}$) may be formed, 
through the Spitzer instability, runaway core collapse 
and merger of the O stars (Portegies Zwart et al.~2004; 
Portegies Zwart \& McMillan 2002; 
Rasio, Freitag \& G\"{u}rkan~2003; G\"{u}rkan, Freitag \& Rasio 2004).
One of the open questions (Portegies Zwart et al.~2004) 
is what range of cluster masses 
and radii offers the best chance for the core collapse/stellar 
coalescence process to occur within the lifetime 
of its more massive O stars ($\approx 3$ Myr). 
%The presence of a ULX in a young cluster with 
%a mass $\approx 3.5 \times 10^5 M_{\odot}$
However, we have found that this ULX in NGC\,4559 
is not located in or near a bright cluster: hence, this formation 
channel seems unlikely for this object, 
unless the parent cluster has already dispersed.
The tidal disruption of a massive parent cluster on a timescale 
$< 30$ Myr is also unlikely\footnote{Clusters with an initial 
(embedded) mass between $\sim$ a few $\times 10^3$ and 
$\sim$ a few $\times 10^4 M_{\odot}$ 
do get disrupted on a shorter timescale because 
of explosive gas losses (Kroupa \& Boily 2002); 
however, in that case the dispersal occurs 
on a timescale $\tau \la t_{\rm cr} \approx 10^{-3} 
t_{\rm rh} \ll t_{\rm cc} \approx 10^{-1} t_{\rm rh}$, 
where $t_{\rm cr}$ is the crossing timescale,
$t_{\rm cc}$ is the core-collapse timescale, and 
$t_{\rm rh}$ is the relaxation timescale (Spitzer 1987). 
In other words, the disruption would occur before 
there is enough time for the core to collapse.  Moreover, 
the mass of the collapsed core is found to be 
$\sim 10^{-3}$ times the total cluster mass 
(G\"{u}rkan et al.~2004); hence, clusters 
in this mass range would probably not be able 
to produce BHs more massive than $\sim 50 M_{\odot}$.}.

On the other hand, low metal abundance may, in principle, 
have an effect at at least two different stages: 
by allowing the formation of more massive progenitor stars; 
and by helping the formation of a more massive compact object 
from a ``normal'' progenitor star.
%and by increasing the mass transfer rate 
%via Roche lobe overflow.
The first of these effects is usually invoked 
to predict the formation of massive Population III stars 
(and, subsequently, of intermediate-mass BH remnants) 
at high redshift (Madau \& Rees 2001). 
For metal abundances $Z \la 10^{-4}$, 
molecular hydrogen is the only effective coolant in 
dense cloud cores. The cloud temperature cannot decrease 
below $\sim$ a few $10^{2}$ K regardless of density, 
and the Jeans mass remains $\sim$ a few $10^{2} M_{\odot}$ 
at all stages. Thus, a dense clump will collapse 
into a single, very massive star (Bromm 2004, 
and references therein) without further sub-fragmentation. 
This simple mechansism does not work for SMC-type metal 
abundances such as those found in NGC\,4559 
and in other nearby metal-poor dwarf galaxies with ULXs. 
It is also highly unlikely that zero-metallicity gas 
could still be present in the outskirts of those galaxies.
However, it is possible that other mechanism 
may lead to the same effect, making the cooling timescale  
longer than the dynamical timescale (timescale for collapse), 
and therefore reducing the fragmentation during the cloud 
collapse. For example, we may speculate that 
a combination of reduced metal abundance and 
external heating of the cloud cores 
during the galaxy collision process 
(through shocks and turbulence) might lead 
to the formation of more massive stars and remnants. 
The balance of heating and cooling in a molecular cloud core 
can be approximately expressed through a polytropic equation 
of state, $P = K\rho^{\gamma}$. It was suggested  
(Spaans \& Silk 2000; Li, Klessen \& Mac Low 2003) 
that a stiff polytropic index $\gamma > 1$ should reduce 
or suppress fragmentation during the molecular core collapse, 
leading to the formation of imore massive stars. 
This naturally occurs in a gas with primordial metal 
abundances, but it may also occur at higher metallicities 
($Z > 0.1 Z_{\odot}$) for example 
if the gas is irradiated by 
an IR background ($T \sim 100$ K; Spaans \& Silk 2000).

The effect of low metal abundance on the evolution 
of normal early-type stars is to reduce the 
mass-loss rate in the radiatively-driven wind 
($\dot{M}_{\rm w} \sim Z^{0.85}$: Vink, 
de Koter \& Lamers~2001; see also Bouret et al.~2003).
This leads to a more massive stellar core, 
which may then collapse into a more massive BH, 
via normal stellar evolution.
Hence, low abundances may explain the formation 
of isolated BHs with masses up to $\approx 50 M_{\odot}$ 
(i.e., up to $\sim$ half of the mass of the progenitor star) 
and isotropic Eddington luminosities 
$\approx 7 \times 10^{39}$ erg s$^{-1}$.
Two-dimensional hydrodynamic simulations 
of core collapse in massive stars by Fryer (1999) and 
Fryer \& Kalogera (2001) found that they could not produce 
BH remnants more massive than $\approx 20 M_{\odot}$.
However, they did not rule out the possibility 
of higher BH masses formed from stellar progenitors 
more massive than $\approx 40 M_{\odot}$ if 
wind losses are negligible. 

\section{Mass of the accreting BH}

Based on the previous qualitative arguments, if 
accreting BHs with masses $\sim 50$--$100 M_{\odot}$ (more 
massive than those found in Milky Way X-ray binaries) 
can indeed be formed from stellar evolution processes, 
they would be natural candidates to explain most ULXs, 
up to isotropic luminosities $\approx 10^{40}$ erg s$^{-1}$, 
without additional assumptions on the geometry of emission.
However, NGC\,4559 X7 has an isotropic luminosity 
persistently $\ga 2 \times 10^{40}$ erg s$^{-1}$ 
in the $0.3$--$10$ keV band, and an extrapolated 
bolometric luminosity $> 5 \times 10^{40}$ erg s$^{-1}$ 
(Paper I): hence, it may still be too bright 
to be a ``normal'' stellar-mass remnant, 
if we assume strictly Eddington-limited isotropic emission.
Therefore, we are left with three alternatives for this source. 
A ``normal'' stellar-mass BH 
($M \la 20 M_{\odot}$) would require either strong, collimated 
beaming (it would effectively be a microblazar) 
or strongly super-Eddington emission.
A $\sim 50$--$100 M_{\odot}$ BH accreting from a $15$--$25 M_{\odot}$ 
supergiant companion would require only mildly anisotropic emission  
(geometrical beaming of a factor $\sim 5$--$10$; King 2004) 
or mildly super-Eddington luminosity. 
An intermediate-mass BH ($M \ga 500 M_{\odot}$) 
would allow for isotropic, sub-Eddington emission.

The X-ray spectral data (Paper I) show the presence 
of a very soft thermal component ($kT_{\rm in} 
\approx 0.12$--$0.15$ keV); if fitted with a standard 
disk-blackbody model truncated at the Schwarzschild 
innermost stable orbit, its luminosity and temperature 
together suggest a mass $\approx 1000 M_{\odot}$.
It has been suggested (King \& Pounds 2003) 
that the very soft thermal component seen in this and 
other ULXs may not be an indicator 
of the inner-disk temperature, but may 
instead be the result of photon down-scattering 
in an optically-thick outflow shrouding 
the central X-ray source. This radiatively-driven 
outflow would be more likely for X-ray sources 
emitting close or above the Eddington limit.
If we accept this argument, we cannot rule out 
the alternative interpretation of this system as a 
$\sim 50$--$100 M_{\odot}$ BH, with Eddington 
or mildly super-Eddington emission and perhaps 
mildly anisotropic emission (gaining a factor $\la 10$ 
in the observed flux from the combination 
of the two effects).  On the other hand, 
Miller, Fabian \& Miller~(2004) argue that the
presence of a power-law hard component of similar 
luminosity ($\sim 10^{40}$ erg s$^{-1}$) to that 
of the soft thermal component requires a BH with mass
$\sim 1000 M_{\odot}$.
The X-ray timing data (Paper I) suggest the presence 
of a break in the power density spectrum at $\approx 0.03$ Hz; 
this can indicate either a BH mass $\approx 40 M_{\odot}$ 
or $\approx 1000 M_{\odot}$, depending on
the identification of the break in comparison 
with standard power density spectra of X-ray binaries 
and AGN. In any case, we can conclude that 
these X-ray findings 
are not consistent with a low-mass, relativistically-beamed 
microblazar scenario, and suggest a BH mass $\ga 50 M_{\odot}$.
(See also Davis \& Mushotzsky (2004) 
for a more general argument against strong beaming.)
%Instead, both the other alternatives are consistent with 
%the X-ray timing and spectral observations.

The optical data indicate an association with a young region 
of massive star formation at the edge of the galactic disk 
(age $< 30$ Myr). They also rule out an association 
between the ULX and massive star clusters; 
hence, they rule out at least one mechanism 
of formation for an intermediate-mass BH.
Other formation processes have been proposed 
for such objects: e.g., from Population III stars. 
However, in this case, rather contrived 
hypotheses would be required to explain its location 
in a young star-forming complex on the galactic 
disk plane, and the capture of a companion star. 
Therefore, on balance, the optical data 
are more consistent with a BH originating  
in the recent episode of massive star formation, 
and accreting from an OB companion.

A possible counterargument against the existence of 
a $\sim 50$--$100 M_{\odot}$ BH in this field is that 
it would require a progenitor star at least twice 
as massive; yet, all the stars we see today are consistent 
with normal OB stars, and there are no extraordinarily 
bright stellar objects. This may be the case if 
the peculiar physical conditions in the molecular gas 
that led to the formation of very massive stars 
were present only for a short time at the beginning 
of the star-forming episode, when the perturbation 
was stronger. All the very massive stars formed 
at the time have already died, and star formation proceeds 
today in a more conventional way, while the density wave 
expands and weakens. Another possibility is that, 
if such massive stars still exist today in this field, 
we may have mistaken them for star clusters. 
For example, we cannot rule out that the bright 
($M_V \approx -9.5$), unresolved 
source described as a cluster in Sect 4.4 might 
instead be a very massive star. In conclusion, we argue that 
a ``young'' $\sim 50$--$100 M_{\odot}$ BH formed 
from stellar evolution processes is consistent  
with both the optical and the X-ray data. 
A BH with a mass $\sim 1000 M_{\odot}$ would also 
be consistent with (and in some respects preferred by) 
the X-ray data, but much more difficult to reconcile 
with the optical results.
Further multi-band studies of this ULX 
(in particular, radio and infrared) are necessary 
to constrain the mass of the BH.

%Metal abundance may also affect the mass transfer rate 
%in a high-mass X-ray binary:
%(for example, metal-poor stars spend a longer fraction of 
%their life as red supergiants) 
%the orbital separation of the binary components increases faster 
%for a higher non-conservative mass transfer; that is, 
%for a higher mass-loss rate from the binary system in a wind.
%For a sufficiently high mass loss, 
%the binary system and hence the Roche lobe of the donor star 
%would expand faster than the expansion of the donor star 
%due to its nuclear evolution; the system would become 
%detached and Roche-lobe overflow would cease, 
%until the donor star regains contact with the Roche lobe. 
%(Podsiadlowski et al.~2003)
%For example, it was suggested (Podsiadlowski et al.~2003) that, 
%in a system such as Cyg X-1, the O-type donor star 
%is not persistently filling its Roche lobe 
%because of the strong mass-loss rate in a wind, 
%which acts to widen the orbit and shrink the secondary.
%Hence, a reduced stellar wind could cause the O-type donor 
%to fill the Roche lobe for a longer time, ensuring 
%a higher mass-accretion rate onto the BH primary, 
%necessary to explain the observed ULX phase.
%Moreover, a reduced mass-loss rate in the winds 
%from the O stars may have an effect on the timescale 
%for cluster evaporation and core collapse, and hence 
%on the likelihood that an intermediate-mass BH 
%can be produced in a young star cluster core via the Spitzer instability.
%A detailed investigation of these effects is beyond 
%the scope of this work.

\section{Conclusions}

We have used {\it HST}/WFPC2 images in the $B$, $V$, $I$ bands, 
an {\it XMM-Newton}/OM image in the UV band, 
and a ground-based H$\alpha$ image to study the star-forming 
complex around a bright ULX at the edge of the late-type spiral NGC\,4559.
We find that star formation in the kpc-size complex 
near the ULX has occurred 
mostly over the last $\approx 30$ Myr, and is still ongoing 
in the southern part of the complex.  The ULX is co-located  
with a small group of OB stars, but is not located 
in the most active region of the star-forming complex.
Also, it does not appear to be associated with any massive 
young clusters or any extraordinary massive star: 
the brightest point source 
in the {\it Chandra} error circle is consistent 
with a single blue supergiant of mass 
$\approx 20 M_{\odot}$. 
A few other stars are resolved inside the error circle: 
mostly blue and red supergiants with inferred masses 
$\approx 10$--$15 M_{\odot}$.
The candidate donor stars near X7 are consistent 
with an age $\approx 20$ Myr, except for the brightest 
one, for which we estimate an age $\approx 10$ Myr 
if it is a single, isolated star. However, it is possible that this 
star appears brighter and younger than its real age because 
of the effect of X-ray irradiation, 
if it is the true optical counterpart.
H$\alpha$ emission is also seen at the position of this star, 
and may come from X-ray ionized gas around the ULX.

Regardless of which is the true counterpart, 
it is very likely that the NGC\,4559 X7 ULX is a young 
system (age $\approx 20$ Myr) with a high-mass donor. 
This is analogous to what is found for the few other 
ULXs with an identified optical counterpart (Liu et al.~2002, 2004; 
Zampieri et al.~2004).
It is also likely (given its X-ray luminosity) that accretion 
occurs via Roche lobe overflow, and that the accreting BH is more 
massive than the secondary star, unlike typical 
high-mass X-ray binaries in the Galaxy. This would imply  
that the orbit widens as mass is transferred, 
in the absence of other mechanisms to remove angular 
momentum from the system. We noted, qualitatively, 
that a supergiant donor may ensure a steady mass transfer 
rate, because its thermal equilibrium radius expands 
as its mass decreases (unlike what happens for main-sequence stars). 
If the supergiant donor is massive enough ($\ga 15 M_{\odot}$), 
it may have a transfer rate $\sim 10^{-5} M_{\odot}$ yr$^{-1}$ 
over its nuclear timescale $\sim 10^6$ yr (in the supergiant phase), 
as required by the observed luminosity.

We cannot conclusively determine whether 
the accreting BH was formed via normal stellar evolution 
(perhaps favoured by the low-metallicity environment 
or by galaxy collision processes), or is instead 
an intermediate-mass BH created via another mechanism. 
However, given the galactic disk location 
and the age of the local stellar population, 
a young BH associated with the recent 
episode of star formation appears much more likely than 
a primordial BH.
Runaway core collapse in a young super star cluster is one mechanism 
proposed for the formation of a young intermediate-mass BH 
in a starburst or star-forming region.
However, our optical study does rule out this scenario for X7.
The only medium-size 
($M \approx 10^{4.2} M_{\odot}$), young ($25\pm5$ Myr) cluster 
in the field is located at $\ga 600$ pc from the ULX; 
this distance also rules out a direct connection 
via BH ejection. Further analysis of the {\it Chandra} data 
(see also Paper I) shows that 
there are no X-ray sources associated with this cluster, 
and in fact no other sources in the whole X7 field, brighter 
than the ACIS-S 12-count detection limit, 
corresponding to $\approx 4 \times 10^{37}$ erg s$^{-1}$ 
for standard spectral models suitable to accreting sources.
%If the cumulative luminosity distribution for ``normal'' 
%high-mass X-ray binaries generally observed 
%in star-forming galaxies were applicable to this star-forming
Based on the results of our X-ray and optical studies, 
we suggest that the X7 ULX is consistent 
with a $\sim 50$--$100 M_{\odot}$ BH recently 
formed from stellar evolution processes, 
and accreting from an OB companion via Roche-lobe 
overflow. 

We have also studied the general properties 
of the stellar population in the large star-forming 
complex around X7. Its stellar population 
is consistent with an average star formation rate 
$\sim 10^{-2}$--$10^{-2.5}$ over the last 30 Myr.
The observed optical colors and the blue-to-red 
supergiant ratio suggest a low metal abundance: 
$0.2 \la Z/Z_{\odot} \la 0.4$ 
(using the Padua tracks), or $0.05 \la Z/Z_{\odot} \la 0.2$ 
(from the Geneva tracks).
This is also consistent with the low abundance found for the 
absorbing medium towards the X7 ULX, from our {\it XMM-Newton} 
study (Table 1, and Paper I).

The star-forming complex has a ring-like appearance; it is 
only faintly connected to the outermost spiral arm of NGC\,4559 
(Fig.~1), and there are no other strongly star-forming 
regions nearby.
This suggests that it is an expanding wave of star formation, 
triggered by an initial density perturbation, 
in a region where the gas was only marginally stable 
to gravitational collapse.
Similar isolated, ring- or shell-like star-forming complexes 
have been seen in the Milky Way and in nearby galaxies. 
A central super-star cluster 
or hypernova could have provided the initial trigger. 
However, in this case we propose that the most likely 
trigger was a collision with a satellite dwarf galaxy punching 
through the gas-rich outer disk of NGC\,4559.
A candidate dwarf galaxy is indeed visible a few arcsec 
north-west of the complex (Fig.~2). 
It is dominated by an old stellar population 
(age $\ga 10^9$ yr) with a few clumps of younger stars 
(age $\sim 10^7$ yr). With a stellar mass $\sim 10^6 M_{\odot}$ 
and (by analogy with similar dwarf galaxies) 
a dark matter mass one order of magnitude larger, 
it would have produced a strong density wave 
in the gas-rich outer disk of NGC\,4559. 
We are planning optical spectroscopic 
studies of this object to determine whether it is indeed 
physically associated to NGC\,4559, as we propose, 
and if so, to estimate the physical parameters of the collision.

We note that this system could be 
a scaled-down version of the Cartwheel system (Gao et al.~2003), 
where bright, short-lived ULXs have been produced 
in the expanding wave of star formation, triggered by 
the galaxy collision (King 2004).
Colliding or tidally-interacting systems, 
low-metallicity environments, and perhaps 
a high rate of clustered star formation 
seem to provide favourable conditions 
for the formation of ULXs. We have briefly 
mentioned some possible effects. 
%leaving a detailed analysis to further work.

Finally, we need to take into account the existence of another 
bright ($L_{\rm x} \ga 10^{40}$ erg s$^{-1}$) ULX in this galaxy, 
NGC\,4559 X10, located in the inner galactic disk 
(Fig.~1). A preliminary investigation of its 
optical environment (Paper I) did not reveal a young 
star-forming region or bright massive stars near X10. 
It is possible that X7 and X10 have different ages and 
have been formed via different physical mechanisms. 
Perhaps this comparison can also shed light on the difference 
between typical ULXs in star-forming and elliptical galaxies.
A detailed study of the similarities and differences 
between the environments of these two ULXs is left 
to further work, which will be based on a new set of 
{\it HST}/ACS observations scheduled for 2005 March 
(including an H$\alpha$ observation).

%Further multi-band studies of the X7 field in this galaxy 
%will be important for understanding the nature 
%of this bright ULX, discriminating between 
%a stellar-mass and an intermediate-mass BH primary. 
%At the same time, they will test our speculation 
%that the star-forming complex was the result 
%of a satellite-galaxy collision with a gas-rich disk.

\section*{Acknowledgments}

We thank Doug Swartz, Helmut Jerjen, Christian Motch 
and Fred Rasio for useful discussions and suggestions.
We also thank the referee (Tim Roberts) for his helpful 
comments. RS would have been supported 
by a Marie Curie fellowship, if its payment had not been 
more than a year overdue.

\end{document}